\documentclass[%
 reprint,
superscriptaddress,
 amsmath,amssymb,
 aps,
 prl,
]{revtex4-1}

\usepackage{graphicx}
\usepackage{dcolumn}
\usepackage{bm}
\usepackage{hyperref}
\hypersetup{
   breaklinks=true,   
   colorlinks=true,   
}

\begin{document}

\title{Mass Measurement of \textsuperscript{27}P to Constrain Type-I X-ray Burst Models and Validate the IMME for the A=27, T=$\frac{3}{2}$ Isospin Quartet}%

\author{I. T. Yandow}%
\email{yandow@frib.msu.edu}
\affiliation{Facility for Rare Isotope Beams, East Lansing, Michigan, 48824, USA}
\affiliation{Department of Physics and Astronomy, Michigan State University, East Lansing, Michigan 48824, USA}
\author{A. Abdullah-Smoot}
\affiliation{Department of Physics, Texas Southern University, Houston, Texas, 77004, USA}
\author{G. Bollen}
\affiliation{Facility for Rare Isotope Beams, East Lansing, Michigan, 48824, USA}
\author{A. Hamaker}
\affiliation{Facility for Rare Isotope Beams, East Lansing, Michigan, 48824, USA}
\affiliation{Department of Physics and Astronomy, Michigan State University, East Lansing, Michigan 48824, USA}
\author {C. R. Nicoloff}
\affiliation{Facility for Rare Isotope Beams, East Lansing, Michigan, 48824, USA}
\affiliation{Department of Physics and Astronomy, Michigan State University, East Lansing, Michigan 48824, USA}
\author{D. Puentes}
\affiliation{Facility for Rare Isotope Beams, East Lansing, Michigan, 48824, USA}
\affiliation{Department of Physics and Astronomy, Michigan State University, East Lansing, Michigan 48824, USA}
\author{M. Redshaw}
\affiliation{Department of Physics, Central Michigan University, Mount Pleasant, Michigan 48824, USA}
\affiliation{Facility for Rare Isotope Beams, East Lansing, Michigan, 48824, USA}
\author{K. Gulyuz}
\affiliation{Department of Physics and Astronomy, Michigan State University, East Lansing, Michigan 48824, USA}
\affiliation{Facility for Rare Isotope Beams, East Lansing, Michigan, 48824, USA}
\author{Z. Meisel}
\affiliation{Department of Physics and Astronomy, Ohio University, Athens, Ohio, 45701, USA}
\affiliation{Edwards Accelerator Laboratory, Ohio University, Athens, Ohio, 45701, USA}
\author{W.-J. Ong}
\affiliation{Lawrence Livermore National Laboratory, Livermore, California, 94550, USA}
\affiliation{Department of Physics and Astronomy, Michigan State University, East Lansing, Michigan 48824, USA}
\author{R. Ringle}
\affiliation{Facility for Rare Isotope Beams, East Lansing, Michigan, 48824, USA}
\author{R. Sandler}
\affiliation{Department of Physics, Central Michigan University, Mount Pleasant, Michigan 48824, USA}
\affiliation{Facility for Rare Isotope Beams, East Lansing, Michigan, 48824, USA}
\author{S. Schwarz}
\affiliation{Facility for Rare Isotope Beams, East Lansing, Michigan, 48824, USA}
\author{C. S. Sumithrarachchi}
\affiliation{Facility for Rare Isotope Beams, East Lansing, Michigan, 48824, USA}
\author{A. A. Valverde}
\affiliation{Physics Division, Argonne National Laboratory, Lemont, Illinois, 60439, USA}
\affiliation{Department of Physics and Astronomy, University of Manitoba, Winnipeg, Manitoba MB R3T 2N2, Canada}
\date{\today}%

\begin{abstract}
\noindent
\begin{description}
\item[Background]
Light curves are the primary observable of type-I x-ray bursts. Computational x-ray burst models must match simulations to observed light curves. Most of the error in simulated curves comes from uncertainties in $rp$ process reaction rates, which can be reduced via precision mass measurements of neutron-deficient isotopes in the $rp$ process path.
\item[Purpose]
Perform a precise atomic mass measurement of $^{27}$P. Use this new measurement to calculate $rp$ process reaction rates and input these rates into an x-ray burst model to reduce simulated light curve uncertainty. Use the mass measurement of $^{27}$P to validate the Isobaric Multiplet Mass Equation (IMME) for the A=27 T=$\frac{3}{2}$ isospin quartet which $^{27}$P belongs to.
\item[Method]
High-precision Penning trap mass spectrometry utilizing the ToF-ICR technique was used to determine the atomic mass of $^{27}$P. The MESA code (Modules for Experiments in Stellar Astrophysics) was then used to simulate x-ray bursts using a 1D multi-zone model to produce updated light curves.
\item[Results]
The mass excess of $^{27}$P was measured to be -670.7(6) keV, a fourteen-fold precision increase over the mass reported in the 2020 Atomic Mass Evaluation (AME2020). The $^{26}$Si($p ,\gamma$)$^{27}$P--$^{27}$P($\gamma,p$)$^{26}$Si rate equilibrium has been determined to a higher precision based on the precision mass measurement of $^{27}$P. X-ray burst light
curves were produced with the MESA code using the new reaction rates. Changes in the mass of $^{27}$P seem to have minimal effect on light curves, even in burster systems tailored to maximize impact. 
\item[Conclusion]
The mass of $^{27}$P does not play a significant role in x-ray burst light curves. It is important to understand that more advanced models do not just provide more precise results, but often qualitatively different ones. This result brings us a step closer to being able to extract stellar parameters from individual x-ray burst observations. In addition, the IMME has been validated for the $A=27, T=3/2$ quartet. The normal quadratic form of the IMME using the latest data yields a  reduced $\chi^2$ of 2.9. The cubic term required to generate an exact fit to the latest data matches theoretical attempts to predict this term.
\end{description}
\end{abstract}

\maketitle

\section{I. Introduction}
\subsection{A. Type-I x-ray bursts}
Type-I x-ray bursts (XRB) are astronomical events that occur in binary star systems with one neutron star and one companion star which has expanded and filled its Roche lobe \cite{RocheLobe}, causing hydrogen and helium-rich material to flow from the companion star to the neutron star.
This accreted material builds up on the surface of the very dense neutron star and is compacted by the extreme gravitational force. The temperature and density of the proton-rich neutron star atmosphere increase continuously as more material is added until fusion and thermonuclear runaway are triggered.  This commences the rapid proton capture ($rp$) process \cite{rpProcess}.

The $rp$ process produces neutron-deficient nuclei lighter than A$\sim$106 via a series of proton captures ($p,\gamma$), photodisintegrations ($\gamma,p$), $\alpha$ induced reactions ($\alpha,p$), and $\beta ^+$-decays \cite{NucleosynthesisInXRB}. The initial thermonuclear flash and $rp$ process rapidly and drastically increase the temperature in the atmosphere of the neutron star, resulting in a sharp increase in x-ray luminosity followed by a slow drop-off---a type-I x-ray burst \cite{GammaRayBursts}. The x-ray luminosity coming from a burst is called the light curve. It is the primary observable from x-ray bursts and must be accurately modeled in order for stellar parameters to be extracted from observed x-ray burst data.

Nuclear data for the nuclei along the $rp$ process reaction pathway are critical for modeling x-ray bursts and their light curves. For some nuclei, slight changes in mass result in a change in the direction of the net ($p,\gamma$)-($\gamma,p$) flow due to the exponential dependence of photodisintegration on the Q-value. This flow change can lead to a significant shift in the energy production and therefore shape of the light curve. 

Of particular importance to modeling the energy production in an x-ray burst is the determination of the intensities of nuclear reactions at ``waiting point nuclei"---nuclei with relatively long half-lives of at least a few seconds. $^{26}$Si is an $rp$ process waiting point nucleus with a $\beta ^+$-decay halflife of 2.25 s. The uncertainty in the flow out of $^{26}$Si was primarily due to the mass uncertainty in $^{27}$P, which is measured in this work.

A sensitivity study by Schatz and Ong \cite{SensitivityStudy} identified $^{27}$P as one of only three nuclei that had a measurable effect on the light curve of a typical hydrogen/helium burst. 
The mass used in the study was the Atomic Mass Evaluation 2016 (AME2016)-reported mass excess of -722.5(26.3) keV \cite{AME2016}. The mass of $^{27}$P is necessary to determine the equilibrium of the $^{26}$Si($p ,\gamma$)$^{27}$P--$^{27}$P($\gamma,p$)$^{26}$Si reaction, which is important to determine the branching between proton capture ($p ,\gamma$) and $\alpha$ capture ($\alpha,p$) on $^{26}$Si. 
The reduction of the burst simulation uncertainties from the mass of $^{27}$P to a negligible level required a mass measurement with an uncertainty of $\sim$1~keV. In this article we present such a measurement performed at the National Superconducting Cyclotron Laboratory (NSCL) \cite{NSCL} and measured by the Low Energy Beam Ion Trap (LEBIT) 9.4~T Penning trap mass spectrometer \cite{LEBIT}.\\

\subsection{B. Isobaric multiplet mass equation}
The isobaric multiplet mass equation (IMME) formalism treats protons and neutrons as degenerate states of the same hadron which are simply different projections of the ``isospin" quantum number: $T$. The free neutron has isospin projection $T_z = +1/2$, and the free proton $T_z = -1/2$. In a nucleus with $A=N+Z$ nucleons, isospin coupling yields $T = (N-Z)/2$, and allows isospin projections $T_z=|T|, |T|+1,...,A/2$. Nuclei with the same number of nucleons, $A$, (i.e. isobars) can have states with the same isospin, $T$, and similar properties. These are called isobaric analog states. Knowledge of isobaric analog states---some of which are excited states---can be used to predict properties of other nuclei in the same isospin-degenerate multiplets.

Perturbation theory can be used to calculate corrections to the mass of the isobaric analog states. When taken to first order this yields the isobaric multiplet mass equation \cite{IMME}:
\begin{equation}
    ME(A,T,T_z) = a(A,T)+b(A,T)T_z+c(A,T)T_z^2
    \label{IMME}
\end{equation}
where $ME$ is the mass excess, $A$ is the mass number, $T$ and $T_z$ are the isospin and its projection, and $a,b,$ and $c$ are coefficients determined theoretically, or by fitting to mass measurements.
Certain nuclear properties---such as second-order Coulomb effects, three-body interactions, and isospin-mixing---require the addition of the terms $dT_z^3$ and $eT_z^4$. The $d$ and $e$ coefficients are expected to have comparatively small magnitudes unless there is a substantial breakdown of isospin symmetry \cite{SurbrookIMME}. Attempts have been made to theoretically explain and predict the cubic $d$ coefficient of the IMME \cite{IMMEDCoef, IMME1, IMME2, IMME3, IMME4}. Predictions have some agreement with experimentally measured \textit{d}-coefficients, but there are several outlier masses that require large, difficult-to-predict \textit{d}-coefficients to properly describe the masses of an isospin multiplet\cite{IMMEDCoef}. In this work we use the precision mass measurement of $^{27}$P performed at LEBIT to evaluate the predictive capabilities of the IMME and determine whether the $A=27$, $T=\frac{3}{2}$ isospin quartet requires a substantial \textit{d}-coefficient in order to accurately describe the masses of the isobaric analog states.





\section{II. Experimental Method and Analysis}
The LEBIT is the only Penning trap mass spectrometry facility able to perform high-precision measurements on rare isotopes produced by projectile fragmentation. In this experiment, short-lived $^{27}$P was generated by impinging 150MeV/u $^{36}$Ar on a 1034mg/cm$^2$ Be target at the Coupled Cyclotron Facility at the NSCL. The beam produced was then sent through the A1900 fragment separator with a 150mg/cm$^2$ 99.99\% pure aluminum wedge \cite{NSCL} to separate the secondary beam.

The beam proceeded to the beam-stopping area \cite{GasCatcher} via a momentum compression beamline, where it was degraded with aluminum degraders of total thickness 2759 $\mu$m before passing through a 4.1 mrad aluminum wedge with center thickness 1016 $\mu$m. The beam entered the gas cell at an energy of less than 1 MeV/u. In the gas cell, ions were stopped in high-purity helium gas at about 52 torr and a temperature of -7$^{\circ}$C. During collisions with the helium gas, the highly charged ions recombined down to the charge state $+1$. The ions were transported through the gas cell by a combination of rf and dc fields and gas flow. They were then extracted into a radio frequency quadrupole (RFQ) ion guide and separated by a magnetic dipole mass separator with a resolving power of approximately 1500.

The activity of the beam after the dipole mass separator was measured with an insertable Si detector. The highest activity was found at a charge-to-mass ratio of $A/Q=43$. This indicated that the majority of the $^{27}$P was being extracted in the form of singly ionized phosphorus-oxide, $^{27}$PO$^+$, though there were trace amounts ($\sim$1\%) of $^{27}$PO$_2$$^+$ detected as well.

By keeping all ion transport electrodes in both the gas stopping facility and the LEBIT laboratory close to 30~kV, but the transport electrodes in between close to ground, the ions accelerated rapidly to LEBIT but were again slowed when they entered LEBIT. Once in LEBIT, the ions entered a helium gas-filled RFQ ion cooler buncher \cite{CoolerBuncher}. The ions were accumulated, stopped in room temperature helium, and released to the LEBIT 9.4~T Penning trap. A fast kicker in the beam line leading from the cooler buncher to the trap was used as a time-of-flight mass separator. It only allowed $^{27}$PO$^+$ and molecular contaminants with a similar mass-to-charge ratio, $A/Q = 43\pm1$, to enter the Penning trap. The isobaric contaminant HCNO$^+$ was used as a calibration ion of well-known mass.

LEBIT's 9.4~T Penning trap is made of a high-precision hyperbolic electrode arrangement in an actively shielded magnet \cite{PenningTrap}. Electrodes leading up to the trap decelerated the ion pulses before they entered. The final section of these electrodes is quadrisected radially, with each segment's voltage independently controllable, to create a Lorentz Steerer \cite{LorentzSteerer}. The Lorentz Steerer controlled how far off-center the ions entered the trap. Once captured, contaminant ions were driven out using dipole cleaning \cite{DipoleCleaning}, which excited the motion of the contaminants using azimuthal rf dipole fields at their reduced cyclotron frequency ($f_+$). Then a mass measurement was performed on the---now isolated---ions of interest.

Penning trap mass measurements do not directly yield mass as a result. Instead, results are given in terms of the frequency ratio
\begin{equation}
    R=\frac{f_{ref}^{int}}{f_c}.
\end{equation}  
Here $f_c$ is the cyclotron frequency of the ion of interest: the frequency at which an ion of charge $q$ and mass $m$ precesses in a uniform magnetic field $B$, given by the equation
\begin{equation}
    f_c=\frac{qB}{m}.
\end{equation}  
$f_{ref}^{int}$ is the time-interpolated cyclotron frequency of a calibration ion with well-known mass. In this measurement, the cyclotron frequency of the calibration ion, HCNO$^+$, was measured before and after each $^{27}$PO$^+$ measurement. The final reported atomic mass $M$ is found by taking the average of all of the frequency ratios, $\bar{R}$, and using
\begin{equation}
    M=\bar{R}[M_{ref}-m_e]+m_e
    \label{eq:MFromRbar}
\end{equation}
where $M_{ref}$ is the atomic mass of the neutral reference atom/molecule---in this case HCNO---and $m_e$ is the electron mass. Electron binding energy is generally neglected, as it is on the order of a few eV's; the dominant statistical uncertainties in this work are two orders of magnitude greater than this.

\begin{figure}[ht]
        \includegraphics[width=\columnwidth]{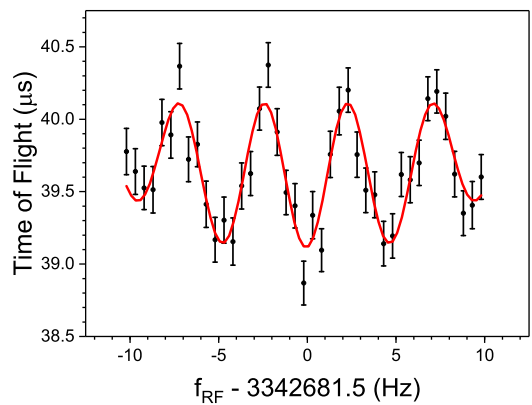}
    \caption{A sample 250~ms total excitation time $^{27}$PO$^+$ time-of-flight ion cyclotron Ramsey resonance measurement. The central dip in the interference pattern corresponds to the resonant cyclotron frequency. The red curve is a fit to the theoretical profile \cite{Ramsey}.}
    \label{fig:RamseyResonance}
\end{figure}

Once the $^{27}$PO$^+$ ions were trapped and cleaned of contaminants, the time-of-flight ion cyclotron resonance technique \cite{ToF1} was used to determine the cyclotron frequency---and therefore mass---of the $^{27}$PO$^+$. A 50-, 100-, 150-, and 200-ms continuous quadrupolar excitation was used to make initial measurements of $^{27}$PO$^+$. The time-of-flight distributions were fit with the theoretical lineshape described in \cite{ToF2}. 
The measured cyclotron frequency was checked against all chemically possible molecules composed of stable or long-lived atoms. It was determined that no potential contaminants were within 3$\sigma$ of the measured resonance, and therefore the observed resonances must be $^{27}$PO$^+$.
Once identified, the excitation was switched to the pulsed Ramsey resonance technique \cite{Ramsey}, which improved precision by a factor of approximately 4. A sample Ramsey resonance of $^{27}$PO$^+$ with a total excitation time of 250~ms can be found in Figure \ref{fig:RamseyResonance}. For both the continuous quadrupolar and Ramsey resonances, in between each $^{27}$PO$^+$ cyclotron frequency measurement, a reference ion measurement of HCNO$^+$ was performed in order to determine the magnetic field strength.

Systematic shifts in $\bar{R}$ (Equation \ref{eq:MFromRbar}) have been found to scale linearly with the mass difference between calibrant and target ions. Systematic shifts can result from trap misalignment with the magnetic field, magnetic field inhomogeneities, and nonharmonic trapping potential imperfections \cite{ToF1}. These mass-dependent shifts have been thoroughly investigated at LEBIT and found to be $\Delta R \sim 2\times 10^{-10}/u$ \cite{MassOffsetError}, which is negligible in comparison to statistical uncertainties when the mass of the reference ion is within a few u of the ion of interest. Because an isobaric reference was used in this measurement, these shifts are certainly negligible.

Further systematic effects include nonlinear temporal shifts in the magnetic field, relativistic effects on $f_c$, and ion-ion interactions in the Penning trap. Nonlinear magnetic field fluctuations have been shown to have an effect less than $1\times10^{-9}$ over one hour \cite{MagneticFieldShift}---which was the approximate duration of each $^{27}$PO$^+$ measurement---making this effect less than statistical uncertainty. Relativistic effects were negligible because of the large ion masses \cite{MagneticFieldShift}. Isobaric contaminants in the trap can lead to systematic frequency shifts. This effect was minimized by performing dipole scans over a broad frequency range in order to detect contaminants. Contaminants were detected by searching for drops in count rate as ions were driven out of the trap. When a contaminant was identified using this method, its reduced cyclotron frequency was added to a list of ``cleaning" frequencies. A dipole excitation for each cleaning frequency was applied, driving out contaminants before the quadrupolar excitation of the ion of interest. For both $^{27}$PO$^+$ and HCNO$^+$, events with six or more detected ions were discarded to avoid potential systematic frequency shifts from Coulomb interactions in the trap.

\section{III. Results}
\begin{figure}[ht]
        \includegraphics[width=\columnwidth]{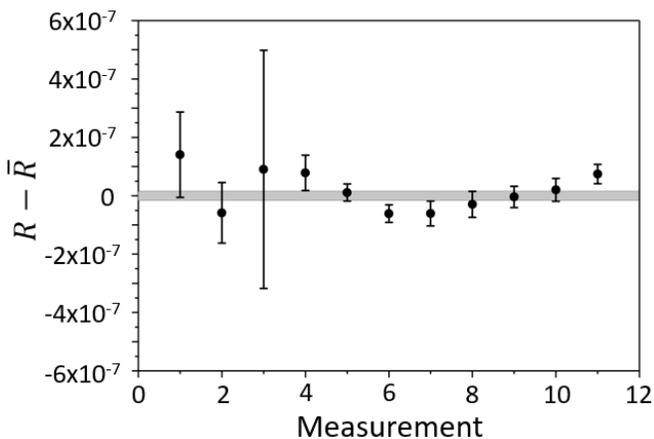}
    \caption{Measured cyclotron frequency ratios $R=\frac{f_{\text{ref}}^{\text{int}}(\text{HCNO}^+)}{f_\text{c}(^{27}\text{PO}^+)}$ relative to the average value $\bar{R} = 1.000270250(16)$. The gray bar represents $\pm 1 \sigma$ uncertainty in $\bar{R}$ and has been scaled by the Birge ratio \cite{BirgeRatio}.}
    \label{fig:P27Results}
\end{figure}

Eleven measurements of $^{27}$PO$^+$ were performed over the course of approximately 15 hours. These resulted in a weighted average of $\bar{R}=1.000270250(16)$, as can be seen in Figure \ref{fig:P27Results}. The individual values of $R$ varied slightly more than would be expected from a Gaussian distribution of measurements. This variation likely derived from a systematic underestimation of uncertainty, resulting in a Birge ratio \cite{BirgeRatio} of 1.23(14). In order to correct this potential error underestimation, the uncertainty of $\bar{R}$ was scaled by the Birge ratio. This correlates to a $^{27}$P mass excess of -670.7(6) keV.\\

\begin{figure}[ht]

        \includegraphics[width=\columnwidth]{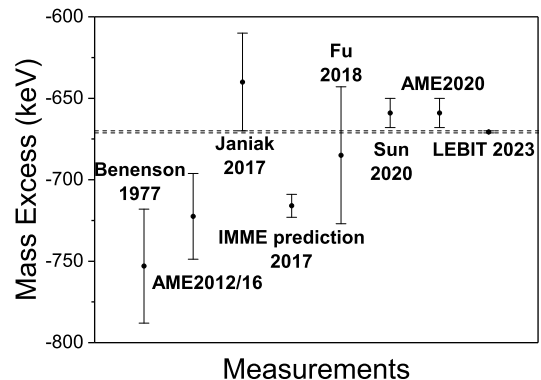}
    \caption{Mass excess of $^{27}$P as measured by LEBIT (dotted lines) compared with recommended values from AME2012/16 \cite{AME2012, AME2016} and AME2020 \cite{AME2020} and values measured by Benenson et al.  \cite{Benenson19771187}, Janiak et al. \cite{PhysRevC.95.034315}, Fu et al. \cite{Fu2018} and Sun et al. \cite{LJSunMeasurement} as well as the value predicted by an IMME calculation \cite{SensitivityStudy}}
    \label{fig:PastP27Measurements}
\end{figure}

Due largely to its astrophysical importance, the mass of $^{27}$P has been measured several times, by Benenson et al. using a split-pole spectrograph \cite{Benenson19771187}, Janiak et al. using $\beta$ delayed proton emission \cite{PhysRevC.95.034315}, Fu et al. using a storage ring \cite{Fu2018}, and Sun et al. using $\beta$ decay spectroscopy of $^{27}$S \cite{LJSunMeasurement}. In addition, Schatz and Ong used the IMME to predict the mass of $^{27}$P in order to improve the precision of x-ray burst simulations\cite{SensitivityStudy}. The LEBIT measurement is shown in comparison to past measurements and predictions in Figure \ref{fig:PastP27Measurements}.
\section{IV. Discussion}
\subsection{A. IMME predictions}
Schatz and Ong used the IMME to predict a $^{27}$P mass excess of -716(7) keV \cite{SensitivityStudy}. They utilized a $^{27}$Si $T=3/2$ isobaric analog state excitation energy of  6626(3) keV \cite{OldSi27} for this calculation. The LEBIT mass measurement found an excess of -670.7(6) keV, a difference exceeding 6$\sigma$. A recent measurement of the $T=3/2$ isobaric analog state of $^{27}$Si performed at Texas A\&M by McCleskey et al. \cite{McCleskey2016} yielded an excitation energy of 6638(1) keV, 12 keV higher than previously measured. Using this new value, a mass excess of -677(4) keV is predicted for $^{27}$P, less than 2$\sigma$ from the LEBIT result. Without the McCleskey measurement, a large cubic term of $d$=8(1) keV would be necessary in order for the IMME to accurately predict the LEBIT value and would indicate a substantial breakdown in isospin symmetry. With the McCleskey excitation energy, the normal quadratic form of the IMME yields an acceptable reduced $\chi^2$ of 2.9. In order to perfectly match measured data, a small \textit{d}-coefficient of $d$=1.1(6) keV must be added, in excellent agreement with the theoretical prediction of this cubic term made by Dong et al. \cite{IMMEDCoef}. While the IMME prediction used in \cite{SensitivityStudy} yielded a poor result for the mass of $^{27}$P, it was missing the critical updated $T=3/2$ isobaric analog state excitation energy of $^{27}$Si. On the one hand, this recommends caution when using the IMME predictively in astrophysics simulations. On the other hand, not only has the IMME been shown to make a reasonable prediction for this multiplet, but the McCleskey measurement \cite{McCleskey2016} and Dong theoretical \textit{d}-coefficient prediction \cite{IMMEDCoef} have been validated. This demonstrates that with accurate data on isobaric analog states, the IMME can be a powerful predictive tool. This result also motivates the revisiting of old excitation energy measurements of isobaric analog states. The most recent measurement of the $T=3/2$ isobaric analog state excitation energy of $^{27}$Si before McCleskey et al. was performed more than half a century ago in 1971 by Barker et al. \cite{OldSi27}. The LEBIT $^{27}$P mass measurement and the latest $A=27, T=3/2$ isobaric analog state information can be found in Table \ref{tab:IMMEFitParams}; the IMME coefficients calculated using this data can be found in Table \ref{tab:UpdatedIMME}.\\

\begin{table}[]
    \centering
    \begin{tabular}{|c|c|c|c|}
        \hline
        Isobar & $T_z$ & ME (keV) & E$_x$ (keV)\\
        \hline
         $^{27}$P & -3/2 & 670.6(6) & 0.0  \\
         $^{27}$Si$^*$ & -1/2 & 12384.5(1) & 6638(1) \\
         $^{27}$Al$^*$ & 1/2 & 17196.86(5) & 6813.8(7)  \\
         $^{27}$Mg & 3/2 & 14586.59(5) & 0.0 \\
         \hline
    \end{tabular}
    \caption{Isobaric analog state information for the $A=27, T=3/2$ quartet. $T_z$ is the isospin projection, ME the mass excess, and E$_x$ the excitation energy. The ground state mass of $^{27}$P is the value presented in this work. The other ground state masses are from AME2020 \cite{AME2020}. Two isobaric analog states are excited states: $^{27}$Si$^*$ \cite{McCleskey2016} and $^{27}$Al$^*$ \cite{Al27}}
    \label{tab:IMMEFitParams}
\end{table}

\begin{table}[]
    \centering
    \begin{tabular}{|c|c|c|}
        \hline
        IMME term & Quadratic Fit (keV) & Cubic Fit (keV)\\
        \hline
         $a$ & 8118.9(6) & 8119.3(7)  \\
         $bT_z$ & 4638.6(2) & 4636(1)  \\
         $cT_z^2$ & -217.9(3) & -218.1(3) \\
         $dT_z^3$ & n/a & 1.1(6) \\
         \hline
    \end{tabular}
    \caption{Regular (quadratic) and cubic IMME coefficient values for the $A=27, T=3/2$ quartet. The IMME is defined as $ME = a+bT_z+cT_z^2(+dT_z^3)$. The reduced $\chi^2$ of the quadratic fit is 2.9. The cubic fit has the same number of fitted parameters as data points---four---so reduced $\chi^2$ is not defined.}
    \label{tab:UpdatedIMME}
\end{table}

\subsection{B. Astrophysical implications}
The LEBIT $^{27}$P mass was used to calculate the proton capture rate on the $^{26}$Si waiting point using the techniques and resonance properties described in Sun et al. Section F \cite{Sun_2019}. The dominant narrow resonance relevant for proton capture at x-ray burst temperatures ($\sim$ 0.1-1.2~GK) is the first excited state $3/2^+$ resonance, with contributions from the second excited state $5/2^+$ resonance several orders of magnitude lower for all x-ray burst temperatures. While direct capture is the dominant pathway for proton capture on $^{26}$Si at temperatures below 0.08 GK, it quickly drops to a negligible level compared to the $3/2^+$ resonance at the high temperatures achieved in an x-ray burst. Often, an improved mass measurement contributes to a better understanding of a proton capture rate by determining the proton separation energy between parent and daughter nuclei. This proton separation energy combined with a measurement of excitation energy determines the resonance energy. Because the dominant $3/2^+$ resonance energy was determined directly via beta-delayed proton measurement in \cite{Sun_2019}, the LEBIT mass measurement impacts neither the resonance energy nor the $^{26}$Si$(p,\gamma)^{27}$P forward rate. Therefore, the effect of the updated $^{27}$P mass lies entirely with the photodisintegration rate, $^{27}$P($\gamma,p$)$^{26}$Si---which depends exponentially on the Q-value.

\begin{figure}[ht]

        \includegraphics[width=\columnwidth]{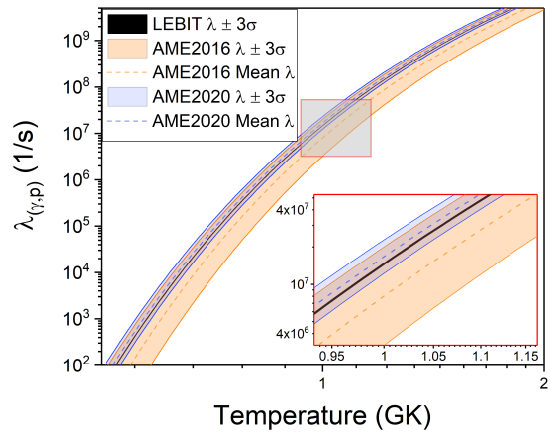}
    \caption{Photodisintegration rate $\lambda_{(\gamma,p)}$ for the reaction $^{27}$P($\gamma,p$)$^{26}$Si using the mass of $^{27}$P  from LEBIT (black), AME2016 \cite{AME2016} (orange), and AME2020 \cite{AME2020} (blue), each varied $\pm3\sigma$ (LEBIT: 1.9 keV, AME2016: 72 keV, AME2020: 27 keV). AME2016 is included to compare to Schatz and Ong's sensitivity study \cite{SensitivityStudy}. A zoomed inset is included to clarify the $\lambda_{(\gamma,p)}$ precision improvement due to the LEBIT measurement.}
    \label{fig:ReverseRates}
\end{figure}

The photodisintegration rate was calculated using the methodologies laid out in Section 3.2 of \cite{ReverseRate}. As can be seen in Figure \ref{fig:ReverseRates}, the LEBIT-based result drastically reduces uncertainty in the photodisintegration rate as compared to that based on AME2016 \cite{AME2016} and AME2020 \cite{AME2020}. The LEBIT rate lies between the rates calculated using these two atomic mass evaluations and is consistent with each of them within $3\sigma$.

Even though this critical rate was much more precisely determined in this work, MESA simulations using the techniques laid out by Z. Meisel in \cite{Meisel_2018} yielded light curves not substantively different than using the AME2016 \cite{AME2016} or AME2020 \cite{AME2020} $^{27}$P mass value. 
In order to determine if the mass of $^{27}$P could have an impact on the light curve of any x-ray burst, the mass accretion rate of the simulated x-ray burster system was decreased compared to the GS 1826-24 clock burster typically used in simulations (decreased from $2.98 \times 10^{-9}$ to $1.23 \times 10^{-9}$ M$_{\odot}$/yr, where M$_{\odot}$ is the mass of the sun, $\sim 2 \times 10^{30}$ kg). This decrease in mass accretion rate increases the burst temperature and helium fraction upon ignition, enhancing the highly temperature-dependent ($\alpha,p$) pathway. The increased ($\alpha,p$)-($p,\gamma$) path competitiveness maximizes the change in the simulated light curve due to the LEBIT $^{27}$P mass measurement or any variation in the $^{27}$P mass. Even in this situation, variations in the mass of $^{27}$P had an inconsequential effect on the light curve.

\begin{figure}[ht]

    \includegraphics[width=\columnwidth]{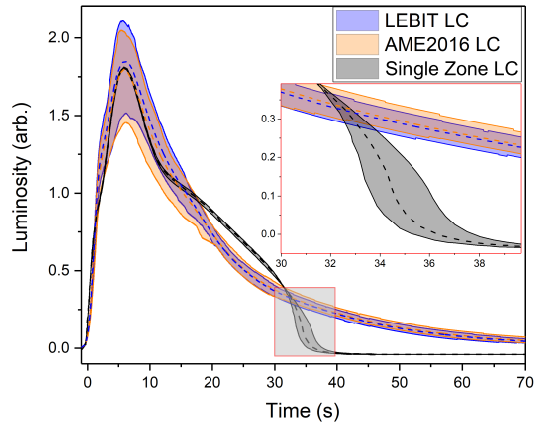}
    \caption{MESA multizone XRB light curve generated using the mass of $^{27}$P from LEBIT (blue) and AME2016 \cite{AME2016} (orange), each varied $\pm3\sigma$ (1.9 keV for LEBIT, 72 keV for AME2016). Their overlap is a dull pink. Single-zone model results from \cite{SensitivityStudy} (gray) are included to show the qualitative difference in light curve tail shape. Burster is ``GS 1826-24"-like, with accretion rate decreased from $2.98 \times 10^{-9}$ to $1.23\times10^{-9}$ M$_{\odot}$/yr to enhance the impact of $^{27}$P mass. AME2020 \cite{AME2020} excluded as it would produce a third curve with near-perfect overlap.}
    \label{fig:MESALightCurve}
\end{figure}

The blue and orange bands and their pink overlapping region in Figure \ref{fig:MESALightCurve} show that the simulated light curve is nearly identical regardless of which mass value for $^{27}$P is used and whether it is varied over the small uncertainty attained by LEBIT, or the uncertainty from AME2016 \cite{AME2016}, which is over forty times as large. The AME2016 recommended mass for $^{27}$P was varied by $\pm3\sigma$ to produce the orange light curve. This mass variation encompasses both the LEBIT and AME2020 \cite{AME2020} mass values of $^{27}$P. It is therefore not surprising that the light curve produced using the LEBIT $^{27}$P mass lies mostly within the uncertainty band of the light curve produced using the AME2016 value. It is surprising that the uncertainty band of the light curve produced using the LEBIT mass is barely reduced. This can be seen by observing that the blue and orange bands in Figure \ref{fig:MESALightCurve} are of comparable magnitude throughout the x-ray burst.

This lack of impact on the simulated light curve from a variation of the mass of $^{27}$P is not what was predicted by the sensitivity study \cite{SensitivityStudy}. The cause of this disparity likely lies in the difference between the simple single-zone XRB model and the complex multi-zone MESA XRB simulation. A single-zone model was used in the sensitivity study in order to be able to approximate the impact of a wide variety of nuclear masses with an achievable amount of computing power. A multi-zone MESA simulation is only feasible to run when varying a small number of parameters such as just the mass of $^{27}$P. One can see from the gray band in Figure \ref{fig:MESALightCurve} that the impact on a single-zone x-ray burst simulation by the mass of $^{27}$P lies primarily in the sharp luminosity drop-off highlighted in the zoomed inset. However, this sudden luminosity drop-off at the end of a light curve does not exist in multi-zone models. As Cyburt et al. point out in \cite{Cyburt_2016}, this region ``is mainly the result of the absence of radiation transport modeling" in single-zone XRB models and that single-zone models run much hotter than multi-zone models. Escape from the $^{26}$Si waiting point via $^{26}$Si($\alpha,p$)$^{29}$P probably requires a temperature higher than is reached in physical systems and multi-zone XRB models in order to compete with proton capture.\\

To test this hypothesis, the flow from the $^{26}$Si waiting point via $^{26}$Si$(p,\gamma)^{27}$P and $^{26}$Si$(\alpha,p)^{29}$P was calculated for an array of temperature and density environments potentially achievable in an XRB. This flow was calculated using a $^{27}$P mass value from AME2016 $\pm$ 3$\sigma$. AME2016 was chosen to demonstrate why the sensitivity study \cite{SensitivityStudy} predicted the mass of $^{27}$P would impact the $rp$ process path and because AME2016 $\pm$ 3$\sigma$ encompasses the LEBIT and AME2020 recommended $^{27}$P masses.
The rate of $^{26}$Si$(\alpha,p)^{29}$P is unmeasured, so 3$\times$ the NON-SMOKER calculated rate---the rate used by the nuclear astrophysics database JINA Reaclib---was chosen. 
The multiplication factor of 3$\times$ was chosen based on an uncertainty study of reaction rates in proton-rich nuclei which found that the true reaction rate is occasionally as much as 3$\times$ the NON-SMOKER rate, and is usually less \cite{NuclearDataUncertainties}. As it was our goal to find the scenario where the $(\alpha,p)$ path was the most competitive, the highest plausible rate was chosen. The $^{27}$P mass-independent path of temperatures and densities in a single-zone XRB model was calculated. Finally, the peak temperature achieved in any MESA simulated burst using any $^{27}$P mass from AME2016, AME2020, or LEBIT was determined.\\

\begin{figure}[ht]
    \centering
    \includegraphics[width=\columnwidth]{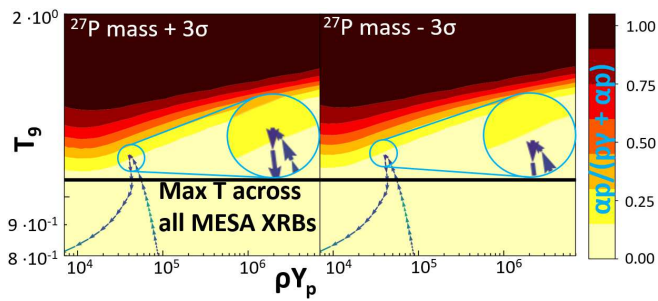}
    \caption{
    Contour plots of the fraction of escape from the $^{26}$Si waiting point via alpha capture---$(\alpha,p)/((p,\gamma)+(\alpha,p))$---during a type-I XRB. Contours generated using a $^{27}$P mass value of AME2016 $\pm$ 3$\sigma$ (left: $+3\sigma$, right : $-3\sigma$). This range is fully inclusive of the LEBIT and AME2020 mass values. T$_9$ is the temperature in GK, $\rho$Y$_p$ is the proton density.\\
    Arrows show the temperature-density profile of the single-zone model over the course of an XRB and are identical in each plot. The horizontal line shows the maximum temperature reached (T $\approx$ 1.1 GK) in any multi-zone MESA XRB simulation using a $^{27}$P mass from AME2016/20, or LEBIT.
    }
    
    \label{fig:WeiJiaCurve}
\end{figure}

The results of these simulations are available in Figure~\ref{fig:WeiJiaCurve}. It shows that the $(\alpha,p)$ path only becomes competitive in the single-zone model when the mass of $^{27}$P is increased by $3\sigma$ based on the AME2016 value (Figure \ref{fig:WeiJiaCurve} left). It is an irrelevant path in a single zone model with the $^{27}$P mass decreased by $3\sigma$ (Figure \ref{fig:WeiJiaCurve} right). Most importantly, it is an irrelevant path in all situations for the multi-zone MESA model (Figure \ref{fig:WeiJiaCurve} horizontal black line). The horizontal black line labeled ``Max T across all MESA XRBs" in Figure \ref{fig:WeiJiaCurve} shows the maximum temperature reached in any MESA simulated x-ray burst with enhanced $(\alpha,p)$ path properties described above and a mass of $^{27}$P from AME2016, AME2020, or LEBIT. This maximum temperature line (T $\approx$ 1.1 GK) never approaches the contour which indicates that $(\alpha,p)$ would begin to compete, even though the $^{26}$Si$(\alpha,p)^{29}$P was set to triple the calculated rate when generating these contours. This shows that the mass of $^{27}$P is critically important for determining whether alpha capture is an escape pathway out of the $^{26}$Si waiting point in single zone XRB models, matching the prediction of \cite{SensitivityStudy}. Multizone models---and most likely real x-ray bursts---never reach a sufficient temperature for the alpha capture pathway to become relevant regardless of the mass of $^{27}$P.\\

\section{V. Conclusions}
$^{27}$P has been measured to over an order of magnitude more precision than prior measurements, with the result ME = -670.7(6) keV. This eliminates the need for further $^{27}$P mass measurements for astrophysical purposes. Its astrophysical impact appears to be less than predicted due to enhanced temperatures reached in single-zone XRB models creating false competition between the proton and $\alpha$ capture pathways out of the $^{26}$Si $rp$ process waiting point. A measurement of the $^{26}$Si($\alpha,p$)$^{29}$P reaction rate would be necessary in order to fully rule out an $\alpha$ capture bypass for high-temperature bursts. This result highlights the importance of following single-zone XRB simulations with multi-zone simulations to validate the impacts predicted by the simpler models. In addition, after some critical updates\cite{McCleskey2016, IMMEDCoef}, the IMME has been validated for the $A=27, T=3/2$ quartet, and the IMME cubic term \textit{d}-coefficient is small enough so as not to cause concern about isospin symmetry breakdown.

\section{Acknowledgements}
This work was conducted with the support of Michigan State University, the US National Science Foundation under contracts nos. PHY-1565546 and PHY-2111185, the DOE, Office of Nuclear Physics under contract no. DE-AC02-06CH11357, DE-SC0015927, and DE-SC0022538, and the Natural Sciences and Engineering Research Council of Canada (NSERC) under Contract No. SAPPJ2018-0028. This work was performed under the auspices of the U.S. Department of Energy by Lawrence Livermore National Laboratory under Contract DE-AC52-07NA27344. Thank you to L.J. Sun for making the author aware of \cite{McCleskey2016}.

\bibliography{refs.bib} 

\begin{thebibliography}{40}%
\makeatletter
\providecommand \@ifxundefined [1]{%
 \@ifx{#1\undefined}
}%
\providecommand \@ifnum [1]{%
 \ifnum #1\expandafter \@firstoftwo
 \else \expandafter \@secondoftwo
 \fi
}%
\providecommand \@ifx [1]{%
 \ifx #1\expandafter \@firstoftwo
 \else \expandafter \@secondoftwo
 \fi
}%
\providecommand \natexlab [1]{#1}%
\providecommand \enquote  [1]{``#1''}%
\providecommand \bibnamefont  [1]{#1}%
\providecommand \bibfnamefont [1]{#1}%
\providecommand \citenamefont [1]{#1}%
\providecommand \href@noop [0]{\@secondoftwo}%
\providecommand \href [0]{\begingroup \@sanitize@url \@href}%
\providecommand \@href[1]{\@@startlink{#1}\@@href}%
\providecommand \@@href[1]{\endgroup#1\@@endlink}%
\providecommand \@sanitize@url [0]{\catcode `\\12\catcode `\$12\catcode
  `\&12\catcode `\#12\catcode `\^12\catcode `\_12\catcode `\%12\relax}%
\providecommand \@@startlink[1]{}%
\providecommand \@@endlink[0]{}%
\providecommand \url  [0]{\begingroup\@sanitize@url \@url }%
\providecommand \@url [1]{\endgroup\@href {#1}{\urlprefix }}%
\providecommand \urlprefix  [0]{URL }%
\providecommand \Eprint [0]{\href }%
\providecommand \doibase [0]{http://dx.doi.org/}%
\providecommand \selectlanguage [0]{\@gobble}%
\providecommand \bibinfo  [0]{\@secondoftwo}%
\providecommand \bibfield  [0]{\@secondoftwo}%
\providecommand \translation [1]{[#1]}%
\providecommand \BibitemOpen [0]{}%
\providecommand \bibitemStop [0]{}%
\providecommand \bibitemNoStop [0]{.\EOS\space}%
\providecommand \EOS [0]{\spacefactor3000\relax}%
\providecommand \BibitemShut  [1]{\csname bibitem#1\endcsname}%
\let\auto@bib@innerbib\@empty
\bibitem [{\citenamefont {Morris}(1994)}]{RocheLobe}%
  \BibitemOpen
  \bibfield  {author} {\bibinfo {author} {\bibfnamefont {S.~L.}\ \bibnamefont
  {Morris}},\ }\href {\doibase 10.1086/133361} {\bibfield  {journal} {\bibinfo
  {journal} {Publications of the Astronomical Society of the Pacific}\ }\textbf
  {\bibinfo {volume} {106}},\ \bibinfo {pages} {154} (\bibinfo {year}
  {1994})}\BibitemShut {NoStop}%
\bibitem [{\citenamefont {{Wallace}}\ and\ \citenamefont
  {{Woosley}}(1981)}]{rpProcess}%
  \BibitemOpen
  \bibfield  {author} {\bibinfo {author} {\bibfnamefont {R.~K.}\ \bibnamefont
  {{Wallace}}}\ and\ \bibinfo {author} {\bibfnamefont {S.~E.}\ \bibnamefont
  {{Woosley}}},\ }\href {\doibase 10.1086/190717} {\bibfield  {journal}
  {\bibinfo  {journal} {The Astrophysical Journal Supplement}\ }\textbf
  {\bibinfo {volume} {45}},\ \bibinfo {pages} {389} (\bibinfo {year}
  {1981})}\BibitemShut {NoStop}%
\bibitem [{\citenamefont {Parikh}\ \emph {et~al.}(2013)\citenamefont {Parikh},
  \citenamefont {José}, \citenamefont {Sala},\ and\ \citenamefont
  {Iliadis}}]{NucleosynthesisInXRB}%
  \BibitemOpen
  \bibfield  {author} {\bibinfo {author} {\bibfnamefont {A.}~\bibnamefont
  {Parikh}}, \bibinfo {author} {\bibfnamefont {J.}~\bibnamefont {José}},
  \bibinfo {author} {\bibfnamefont {G.}~\bibnamefont {Sala}}, \ and\ \bibinfo
  {author} {\bibfnamefont {C.}~\bibnamefont {Iliadis}},\ }\href {\doibase
  https://doi.org/10.1016/j.ppnp.2012.11.002} {\bibfield  {journal} {\bibinfo
  {journal} {Progress in Particle and Nuclear Physics}\ }\textbf {\bibinfo
  {volume} {69}},\ \bibinfo {pages} {225} (\bibinfo {year} {2013})}\BibitemShut
  {NoStop}%
\bibitem [{\citenamefont {Woosley}\ and\ \citenamefont
  {Taam}(1976)}]{GammaRayBursts}%
  \BibitemOpen
  \bibfield  {author} {\bibinfo {author} {\bibfnamefont {S.}~\bibnamefont
  {Woosley}}\ and\ \bibinfo {author} {\bibfnamefont {R.~E.}\ \bibnamefont
  {Taam}},\ }\href {\doibase 10.1038/263101a0} {\bibfield  {journal} {\bibinfo
  {journal} {Nature}\ }\textbf {\bibinfo {volume} {263}},\ \bibinfo {pages}
  {101} (\bibinfo {year} {1976})}\BibitemShut {NoStop}%
\bibitem [{\citenamefont {Schatz}\ and\ \citenamefont
  {Ong}(2017)}]{SensitivityStudy}%
  \BibitemOpen
  \bibfield  {author} {\bibinfo {author} {\bibfnamefont {H.}~\bibnamefont
  {Schatz}}\ and\ \bibinfo {author} {\bibfnamefont {W.-J.}\ \bibnamefont
  {Ong}},\ }\href {\doibase 10.3847/1538-4357/aa7de9} {\bibfield  {journal}
  {\bibinfo  {journal} {The Astrophysical Journal}\ }\textbf {\bibinfo {volume}
  {844}},\ \bibinfo {pages} {139} (\bibinfo {year} {2017})}\BibitemShut
  {NoStop}%
\bibitem [{\citenamefont {Wang}\ \emph {et~al.}(2017)\citenamefont {Wang},
  \citenamefont {Audi}, \citenamefont {Kondev}, \citenamefont {Huang},
  \citenamefont {Naimi},\ and\ \citenamefont {Xing}}]{AME2016}%
  \BibitemOpen
  \bibfield  {author} {\bibinfo {author} {\bibfnamefont {M.}~\bibnamefont
  {Wang}}, \bibinfo {author} {\bibfnamefont {G.}~\bibnamefont {Audi}}, \bibinfo
  {author} {\bibfnamefont {F.~G.}\ \bibnamefont {Kondev}}, \bibinfo {author}
  {\bibfnamefont {W.}~\bibnamefont {Huang}}, \bibinfo {author} {\bibfnamefont
  {S.}~\bibnamefont {Naimi}}, \ and\ \bibinfo {author} {\bibfnamefont
  {X.}~\bibnamefont {Xing}},\ }\href {\doibase 10.1088/1674-1137/41/3/030003}
  {\bibfield  {journal} {\bibinfo  {journal} {Chinese Physics C}\ }\textbf
  {\bibinfo {volume} {41}},\ \bibinfo {pages} {030003} (\bibinfo {year}
  {2017})}\BibitemShut {NoStop}%
\bibitem [{\citenamefont {Morrissey}\ \emph {et~al.}(2003)\citenamefont
  {Morrissey}, \citenamefont {Sherrill}, \citenamefont {Steiner}, \citenamefont
  {Stolz},\ and\ \citenamefont {Wiedenhoever}}]{NSCL}%
  \BibitemOpen
  \bibfield  {author} {\bibinfo {author} {\bibfnamefont {D.}~\bibnamefont
  {Morrissey}}, \bibinfo {author} {\bibfnamefont {B.}~\bibnamefont {Sherrill}},
  \bibinfo {author} {\bibfnamefont {M.}~\bibnamefont {Steiner}}, \bibinfo
  {author} {\bibfnamefont {A.}~\bibnamefont {Stolz}}, \ and\ \bibinfo {author}
  {\bibfnamefont {I.}~\bibnamefont {Wiedenhoever}},\ }\href {\doibase
  https://doi.org/10.1016/S0168-583X(02)01895-5} {\bibfield  {journal}
  {\bibinfo  {journal} {Nuclear Instruments and Methods in Physics Research
  Section B: Beam Interactions with Materials and Atoms}\ }\textbf {\bibinfo
  {volume} {204}},\ \bibinfo {pages} {90} (\bibinfo {year} {2003})},\ \bibinfo
  {note} {14th International Conference on Electromagnetic Isotope Separators
  and Techniques Related to their Applications}\BibitemShut {NoStop}%
\bibitem [{\citenamefont {Ringle}\ \emph {et~al.}(2013)\citenamefont {Ringle},
  \citenamefont {Schwarz},\ and\ \citenamefont {Bollen}}]{LEBIT}%
  \BibitemOpen
  \bibfield  {author} {\bibinfo {author} {\bibfnamefont {R.}~\bibnamefont
  {Ringle}}, \bibinfo {author} {\bibfnamefont {S.}~\bibnamefont {Schwarz}}, \
  and\ \bibinfo {author} {\bibfnamefont {G.}~\bibnamefont {Bollen}},\ }\href
  {\doibase https://doi.org/10.1016/j.ijms.2013.04.001} {\bibfield  {journal}
  {\bibinfo  {journal} {International Journal of Mass Spectrometry}\ }\textbf
  {\bibinfo {volume} {349-350}},\ \bibinfo {pages} {87} (\bibinfo {year}
  {2013})},\ \bibinfo {note} {100 years of Mass Spectrometry}\BibitemShut
  {NoStop}%
\bibitem [{\citenamefont {Weinberg}\ and\ \citenamefont
  {Treiman}(1959)}]{IMME}%
  \BibitemOpen
  \bibfield  {author} {\bibinfo {author} {\bibfnamefont {S.}~\bibnamefont
  {Weinberg}}\ and\ \bibinfo {author} {\bibfnamefont {S.~B.}\ \bibnamefont
  {Treiman}},\ }\href {\doibase 10.1103/PhysRev.116.465} {\bibfield  {journal}
  {\bibinfo  {journal} {Phys. Rev.}\ }\textbf {\bibinfo {volume} {116}},\
  \bibinfo {pages} {465} (\bibinfo {year} {1959})}\BibitemShut {NoStop}%
\bibitem [{\citenamefont {Surbrook}\ \emph {et~al.}(2021)\citenamefont
  {Surbrook}, \citenamefont {Bollen}, \citenamefont {Brodeur}, \citenamefont
  {Hamaker}, \citenamefont {P\'erez-Loureiro}, \citenamefont {Puentes},
  \citenamefont {Nicoloff}, \citenamefont {Redshaw}, \citenamefont {Ringle},
  \citenamefont {Schwarz}, \citenamefont {Sumithrarachchi}, \citenamefont
  {Sun}, \citenamefont {Valverde}, \citenamefont {Villari}, \citenamefont
  {Wrede},\ and\ \citenamefont {Yandow}}]{SurbrookIMME}%
  \BibitemOpen
  \bibfield  {author} {\bibinfo {author} {\bibfnamefont {J.}~\bibnamefont
  {Surbrook}}, \bibinfo {author} {\bibfnamefont {G.}~\bibnamefont {Bollen}},
  \bibinfo {author} {\bibfnamefont {M.}~\bibnamefont {Brodeur}}, \bibinfo
  {author} {\bibfnamefont {A.}~\bibnamefont {Hamaker}}, \bibinfo {author}
  {\bibfnamefont {D.}~\bibnamefont {P\'erez-Loureiro}}, \bibinfo {author}
  {\bibfnamefont {D.}~\bibnamefont {Puentes}}, \bibinfo {author} {\bibfnamefont
  {C.}~\bibnamefont {Nicoloff}}, \bibinfo {author} {\bibfnamefont
  {M.}~\bibnamefont {Redshaw}}, \bibinfo {author} {\bibfnamefont
  {R.}~\bibnamefont {Ringle}}, \bibinfo {author} {\bibfnamefont
  {S.}~\bibnamefont {Schwarz}}, \bibinfo {author} {\bibfnamefont {C.~S.}\
  \bibnamefont {Sumithrarachchi}}, \bibinfo {author} {\bibfnamefont {L.~J.}\
  \bibnamefont {Sun}}, \bibinfo {author} {\bibfnamefont {A.~A.}\ \bibnamefont
  {Valverde}}, \bibinfo {author} {\bibfnamefont {A.~C.~C.}\ \bibnamefont
  {Villari}}, \bibinfo {author} {\bibfnamefont {C.}~\bibnamefont {Wrede}}, \
  and\ \bibinfo {author} {\bibfnamefont {I.~T.}\ \bibnamefont {Yandow}},\
  }\href {\doibase 10.1103/PhysRevC.103.014323} {\bibfield  {journal} {\bibinfo
   {journal} {Phys. Rev. C}\ }\textbf {\bibinfo {volume} {103}},\ \bibinfo
  {pages} {014323} (\bibinfo {year} {2021})}\BibitemShut {NoStop}%
\bibitem [{\citenamefont {Dong}\ \emph {et~al.}(2019)\citenamefont {Dong},
  \citenamefont {Gu}, \citenamefont {Zhang}, \citenamefont {Zuo}, \citenamefont
  {Wang}, \citenamefont {Litvinov},\ and\ \citenamefont {Sun}}]{IMMEDCoef}%
  \BibitemOpen
  \bibfield  {author} {\bibinfo {author} {\bibfnamefont {J.~M.}\ \bibnamefont
  {Dong}}, \bibinfo {author} {\bibfnamefont {J.~Z.}\ \bibnamefont {Gu}},
  \bibinfo {author} {\bibfnamefont {Y.~H.}\ \bibnamefont {Zhang}}, \bibinfo
  {author} {\bibfnamefont {W.}~\bibnamefont {Zuo}}, \bibinfo {author}
  {\bibfnamefont {L.~J.}\ \bibnamefont {Wang}}, \bibinfo {author}
  {\bibfnamefont {Y.~A.}\ \bibnamefont {Litvinov}}, \ and\ \bibinfo {author}
  {\bibfnamefont {Y.}~\bibnamefont {Sun}},\ }\href {\doibase
  10.1103/PhysRevC.99.014319} {\bibfield  {journal} {\bibinfo  {journal} {Phys.
  Rev. C}\ }\textbf {\bibinfo {volume} {99}},\ \bibinfo {pages} {014319}
  (\bibinfo {year} {2019})}\BibitemShut {NoStop}%
\bibitem [{\citenamefont {Brodeur}\ \emph {et~al.}(2012)\citenamefont
  {Brodeur}, \citenamefont {Brunner}, \citenamefont {Ettenauer}, \citenamefont
  {Lapierre}, \citenamefont {Ringle}, \citenamefont {Brown}, \citenamefont
  {Lunney},\ and\ \citenamefont {Dilling}}]{IMME1}%
  \BibitemOpen
  \bibfield  {author} {\bibinfo {author} {\bibfnamefont {M.}~\bibnamefont
  {Brodeur}}, \bibinfo {author} {\bibfnamefont {T.}~\bibnamefont {Brunner}},
  \bibinfo {author} {\bibfnamefont {S.}~\bibnamefont {Ettenauer}}, \bibinfo
  {author} {\bibfnamefont {A.}~\bibnamefont {Lapierre}}, \bibinfo {author}
  {\bibfnamefont {R.}~\bibnamefont {Ringle}}, \bibinfo {author} {\bibfnamefont
  {B.~A.}\ \bibnamefont {Brown}}, \bibinfo {author} {\bibfnamefont
  {D.}~\bibnamefont {Lunney}}, \ and\ \bibinfo {author} {\bibfnamefont
  {J.}~\bibnamefont {Dilling}},\ }\href {\doibase
  10.1103/PhysRevLett.108.212501} {\bibfield  {journal} {\bibinfo  {journal}
  {Phys. Rev. Lett.}\ }\textbf {\bibinfo {volume} {108}},\ \bibinfo {pages}
  {212501} (\bibinfo {year} {2012})}\BibitemShut {NoStop}%
\bibitem [{\citenamefont {Signoracci}\ and\ \citenamefont
  {Brown}(2011)}]{IMME2}%
  \BibitemOpen
  \bibfield  {author} {\bibinfo {author} {\bibfnamefont {A.}~\bibnamefont
  {Signoracci}}\ and\ \bibinfo {author} {\bibfnamefont {B.~A.}\ \bibnamefont
  {Brown}},\ }\href {\doibase 10.1103/PhysRevC.84.031301} {\bibfield  {journal}
  {\bibinfo  {journal} {Phys. Rev. C}\ }\textbf {\bibinfo {volume} {84}},\
  \bibinfo {pages} {031301} (\bibinfo {year} {2011})}\BibitemShut {NoStop}%
\bibitem [{\citenamefont {Henley}\ and\ \citenamefont {Lacy}(1969)}]{IMME3}%
  \BibitemOpen
  \bibfield  {author} {\bibinfo {author} {\bibfnamefont {E.~M.}\ \bibnamefont
  {Henley}}\ and\ \bibinfo {author} {\bibfnamefont {C.~E.}\ \bibnamefont
  {Lacy}},\ }\href {\doibase 10.1103/PhysRev.184.1228} {\bibfield  {journal}
  {\bibinfo  {journal} {Phys. Rev.}\ }\textbf {\bibinfo {volume} {184}},\
  \bibinfo {pages} {1228} (\bibinfo {year} {1969})}\BibitemShut {NoStop}%
\bibitem [{\citenamefont {Bertsch}\ and\ \citenamefont {Kahana}(1970)}]{IMME4}%
  \BibitemOpen
  \bibfield  {author} {\bibinfo {author} {\bibfnamefont {G.}~\bibnamefont
  {Bertsch}}\ and\ \bibinfo {author} {\bibfnamefont {S.}~\bibnamefont
  {Kahana}},\ }\href {\doibase https://doi.org/10.1016/0370-2693(70)90568-X}
  {\bibfield  {journal} {\bibinfo  {journal} {Physics Letters B}\ }\textbf
  {\bibinfo {volume} {33}},\ \bibinfo {pages} {193} (\bibinfo {year}
  {1970})}\BibitemShut {NoStop}%
\bibitem [{\citenamefont {Sumithrarachchi}\ \emph {et~al.}(2020)\citenamefont
  {Sumithrarachchi}, \citenamefont {Morrissey}, \citenamefont {Schwarz},
  \citenamefont {Lund}, \citenamefont {Bollen}, \citenamefont {Ringle},
  \citenamefont {Savard},\ and\ \citenamefont {Villari}}]{GasCatcher}%
  \BibitemOpen
  \bibfield  {author} {\bibinfo {author} {\bibfnamefont {C.}~\bibnamefont
  {Sumithrarachchi}}, \bibinfo {author} {\bibfnamefont {D.}~\bibnamefont
  {Morrissey}}, \bibinfo {author} {\bibfnamefont {S.}~\bibnamefont {Schwarz}},
  \bibinfo {author} {\bibfnamefont {K.}~\bibnamefont {Lund}}, \bibinfo {author}
  {\bibfnamefont {G.}~\bibnamefont {Bollen}}, \bibinfo {author} {\bibfnamefont
  {R.}~\bibnamefont {Ringle}}, \bibinfo {author} {\bibfnamefont
  {G.}~\bibnamefont {Savard}}, \ and\ \bibinfo {author} {\bibfnamefont
  {A.}~\bibnamefont {Villari}},\ }\href {\doibase
  https://doi.org/10.1016/j.nimb.2019.04.077} {\bibfield  {journal} {\bibinfo
  {journal} {Nuclear Instruments and Methods in Physics Research Section B:
  Beam Interactions with Materials and Atoms}\ }\textbf {\bibinfo {volume}
  {463}},\ \bibinfo {pages} {305} (\bibinfo {year} {2020})}\BibitemShut
  {NoStop}%
\bibitem [{\citenamefont {Schwarz}\ \emph {et~al.}(2016)\citenamefont
  {Schwarz}, \citenamefont {Bollen}, \citenamefont {Ringle}, \citenamefont
  {Savory},\ and\ \citenamefont {Schury}}]{CoolerBuncher}%
  \BibitemOpen
  \bibfield  {author} {\bibinfo {author} {\bibfnamefont {S.}~\bibnamefont
  {Schwarz}}, \bibinfo {author} {\bibfnamefont {G.}~\bibnamefont {Bollen}},
  \bibinfo {author} {\bibfnamefont {R.}~\bibnamefont {Ringle}}, \bibinfo
  {author} {\bibfnamefont {J.}~\bibnamefont {Savory}}, \ and\ \bibinfo {author}
  {\bibfnamefont {P.}~\bibnamefont {Schury}},\ }\href {\doibase
  https://doi.org/10.1016/j.nima.2016.01.078} {\bibfield  {journal} {\bibinfo
  {journal} {Nuclear Instruments and Methods in Physics Research Section A:
  Accelerators, Spectrometers, Detectors and Associated Equipment}\ }\textbf
  {\bibinfo {volume} {816}},\ \bibinfo {pages} {131} (\bibinfo {year}
  {2016})}\BibitemShut {NoStop}%
\bibitem [{\citenamefont {Ringle}\ \emph {et~al.}(2009)\citenamefont {Ringle},
  \citenamefont {Bollen}, \citenamefont {Prinke}, \citenamefont {Savory},
  \citenamefont {Schury}, \citenamefont {Schwarz},\ and\ \citenamefont
  {Sun}}]{PenningTrap}%
  \BibitemOpen
  \bibfield  {author} {\bibinfo {author} {\bibfnamefont {R.}~\bibnamefont
  {Ringle}}, \bibinfo {author} {\bibfnamefont {G.}~\bibnamefont {Bollen}},
  \bibinfo {author} {\bibfnamefont {A.}~\bibnamefont {Prinke}}, \bibinfo
  {author} {\bibfnamefont {J.}~\bibnamefont {Savory}}, \bibinfo {author}
  {\bibfnamefont {P.}~\bibnamefont {Schury}}, \bibinfo {author} {\bibfnamefont
  {S.}~\bibnamefont {Schwarz}}, \ and\ \bibinfo {author} {\bibfnamefont
  {T.}~\bibnamefont {Sun}},\ }\href {\doibase
  https://doi.org/10.1016/j.nima.2009.03.207} {\bibfield  {journal} {\bibinfo
  {journal} {Nuclear Instruments and Methods in Physics Research Section A:
  Accelerators, Spectrometers, Detectors and Associated Equipment}\ }\textbf
  {\bibinfo {volume} {604}},\ \bibinfo {pages} {536} (\bibinfo {year}
  {2009})}\BibitemShut {NoStop}%
\bibitem [{\citenamefont {Ringle}\ \emph
  {et~al.}(2007{\natexlab{a}})\citenamefont {Ringle}, \citenamefont {Bollen},
  \citenamefont {Prinke}, \citenamefont {Savory}, \citenamefont {Schury},
  \citenamefont {Schwarz},\ and\ \citenamefont {Sun}}]{LorentzSteerer}%
  \BibitemOpen
  \bibfield  {author} {\bibinfo {author} {\bibfnamefont {R.}~\bibnamefont
  {Ringle}}, \bibinfo {author} {\bibfnamefont {G.}~\bibnamefont {Bollen}},
  \bibinfo {author} {\bibfnamefont {A.}~\bibnamefont {Prinke}}, \bibinfo
  {author} {\bibfnamefont {J.}~\bibnamefont {Savory}}, \bibinfo {author}
  {\bibfnamefont {P.}~\bibnamefont {Schury}}, \bibinfo {author} {\bibfnamefont
  {S.}~\bibnamefont {Schwarz}}, \ and\ \bibinfo {author} {\bibfnamefont
  {T.}~\bibnamefont {Sun}},\ }\href {\doibase
  https://doi.org/10.1016/j.ijms.2006.12.008} {\bibfield  {journal} {\bibinfo
  {journal} {International Journal of Mass Spectrometry}\ }\textbf {\bibinfo
  {volume} {263}},\ \bibinfo {pages} {38} (\bibinfo {year}
  {2007}{\natexlab{a}})}\BibitemShut {NoStop}%
\bibitem [{\citenamefont {Blaum}\ \emph {et~al.}(2004)\citenamefont {Blaum},
  \citenamefont {Beck}, \citenamefont {Bollen}, \citenamefont {Delahaye},
  \citenamefont {Guénaut}, \citenamefont {Herfurth}, \citenamefont
  {Kellerbauer}, \citenamefont {Kluge}, \citenamefont {Lunney}, \citenamefont
  {Schwarz}, \citenamefont {Schweikhard},\ and\ \citenamefont
  {Yazidjian}}]{DipoleCleaning}%
  \BibitemOpen
  \bibfield  {author} {\bibinfo {author} {\bibfnamefont {K.}~\bibnamefont
  {Blaum}}, \bibinfo {author} {\bibfnamefont {D.}~\bibnamefont {Beck}},
  \bibinfo {author} {\bibfnamefont {G.}~\bibnamefont {Bollen}}, \bibinfo
  {author} {\bibfnamefont {P.}~\bibnamefont {Delahaye}}, \bibinfo {author}
  {\bibfnamefont {C.}~\bibnamefont {Guénaut}}, \bibinfo {author}
  {\bibfnamefont {F.}~\bibnamefont {Herfurth}}, \bibinfo {author}
  {\bibfnamefont {A.}~\bibnamefont {Kellerbauer}}, \bibinfo {author}
  {\bibfnamefont {H.-J.}\ \bibnamefont {Kluge}}, \bibinfo {author}
  {\bibfnamefont {D.}~\bibnamefont {Lunney}}, \bibinfo {author} {\bibfnamefont
  {S.}~\bibnamefont {Schwarz}}, \bibinfo {author} {\bibfnamefont
  {L.}~\bibnamefont {Schweikhard}}, \ and\ \bibinfo {author} {\bibfnamefont
  {C.}~\bibnamefont {Yazidjian}},\ }\href {\doibase 10.1209/epl/i2004-10089-5}
  {\bibfield  {journal} {\bibinfo  {journal} {Europhysics Letters}\ }\textbf
  {\bibinfo {volume} {67}},\ \bibinfo {pages} {586} (\bibinfo {year}
  {2004})}\BibitemShut {NoStop}%
\bibitem [{\citenamefont {Bollen}\ \emph {et~al.}(1990)\citenamefont {Bollen},
  \citenamefont {Moore}, \citenamefont {Savard},\ and\ \citenamefont
  {Stolzenberg}}]{ToF1}%
  \BibitemOpen
  \bibfield  {author} {\bibinfo {author} {\bibfnamefont {G.}~\bibnamefont
  {Bollen}}, \bibinfo {author} {\bibfnamefont {R.~B.}\ \bibnamefont {Moore}},
  \bibinfo {author} {\bibfnamefont {G.}~\bibnamefont {Savard}}, \ and\ \bibinfo
  {author} {\bibfnamefont {H.}~\bibnamefont {Stolzenberg}},\ }\href {\doibase
  10.1063/1.346185} {\bibfield  {journal} {\bibinfo  {journal} {Journal of
  Applied Physics}\ }\textbf {\bibinfo {volume} {68}},\ \bibinfo {pages} {4355}
  (\bibinfo {year} {1990})},\ \Eprint
  {http://arxiv.org/abs/https://doi.org/10.1063/1.346185}
  {https://doi.org/10.1063/1.346185} \BibitemShut {NoStop}%
\bibitem [{\citenamefont {K\"onig}\ \emph {et~al.}(1995)\citenamefont
  {K\"onig}, \citenamefont {Bollen}, \citenamefont {Kluge}, \citenamefont
  {Otto},\ and\ \citenamefont {Szerypo}}]{ToF2}%
  \BibitemOpen
  \bibfield  {author} {\bibinfo {author} {\bibfnamefont {M.}~\bibnamefont
  {K\"onig}}, \bibinfo {author} {\bibfnamefont {G.}~\bibnamefont {Bollen}},
  \bibinfo {author} {\bibfnamefont {H.-J.}\ \bibnamefont {Kluge}}, \bibinfo
  {author} {\bibfnamefont {T.}~\bibnamefont {Otto}}, \ and\ \bibinfo {author}
  {\bibfnamefont {J.}~\bibnamefont {Szerypo}},\ }\href {\doibase
  https://doi.org/10.1016/0168-1176(95)04146-C} {\bibfield  {journal} {\bibinfo
   {journal} {International Journal of Mass Spectrometry and Ion Processes}\
  }\textbf {\bibinfo {volume} {142}},\ \bibinfo {pages} {95} (\bibinfo {year}
  {1995})}\BibitemShut {NoStop}%
\bibitem [{\citenamefont {Kretzschmar}(2007)}]{Ramsey}%
  \BibitemOpen
  \bibfield  {author} {\bibinfo {author} {\bibfnamefont {M.}~\bibnamefont
  {Kretzschmar}},\ }\href {\doibase https://doi.org/10.1016/j.ijms.2007.04.002}
  {\bibfield  {journal} {\bibinfo  {journal} {International Journal of Mass
  Spectrometry}\ }\textbf {\bibinfo {volume} {264}},\ \bibinfo {pages} {122}
  (\bibinfo {year} {2007})}\BibitemShut {NoStop}%
\bibitem [{\citenamefont {Gulyuz}\ \emph {et~al.}(2015)\citenamefont {Gulyuz},
  \citenamefont {Ariche}, \citenamefont {Bollen}, \citenamefont {Bustabad},
  \citenamefont {Eibach}, \citenamefont {Izzo}, \citenamefont {Novario},
  \citenamefont {Redshaw}, \citenamefont {Ringle}, \citenamefont {Sandler},
  \citenamefont {Schwarz},\ and\ \citenamefont {Valverde}}]{MassOffsetError}%
  \BibitemOpen
  \bibfield  {author} {\bibinfo {author} {\bibfnamefont {K.}~\bibnamefont
  {Gulyuz}}, \bibinfo {author} {\bibfnamefont {J.}~\bibnamefont {Ariche}},
  \bibinfo {author} {\bibfnamefont {G.}~\bibnamefont {Bollen}}, \bibinfo
  {author} {\bibfnamefont {S.}~\bibnamefont {Bustabad}}, \bibinfo {author}
  {\bibfnamefont {M.}~\bibnamefont {Eibach}}, \bibinfo {author} {\bibfnamefont
  {C.}~\bibnamefont {Izzo}}, \bibinfo {author} {\bibfnamefont {S.~J.}\
  \bibnamefont {Novario}}, \bibinfo {author} {\bibfnamefont {M.}~\bibnamefont
  {Redshaw}}, \bibinfo {author} {\bibfnamefont {R.}~\bibnamefont {Ringle}},
  \bibinfo {author} {\bibfnamefont {R.}~\bibnamefont {Sandler}}, \bibinfo
  {author} {\bibfnamefont {S.}~\bibnamefont {Schwarz}}, \ and\ \bibinfo
  {author} {\bibfnamefont {A.~A.}\ \bibnamefont {Valverde}},\ }\href {\doibase
  10.1103/PhysRevC.91.055501} {\bibfield  {journal} {\bibinfo  {journal} {Phys.
  Rev. C}\ }\textbf {\bibinfo {volume} {91}},\ \bibinfo {pages} {055501}
  (\bibinfo {year} {2015})}\BibitemShut {NoStop}%
\bibitem [{\citenamefont {Ringle}\ \emph
  {et~al.}(2007{\natexlab{b}})\citenamefont {Ringle}, \citenamefont {Sun},
  \citenamefont {Bollen}, \citenamefont {Davies}, \citenamefont {Facina},
  \citenamefont {Huikari}, \citenamefont {Kwan}, \citenamefont {Morrissey},
  \citenamefont {Prinke}, \citenamefont {Savory}, \citenamefont {Schury},
  \citenamefont {Schwarz},\ and\ \citenamefont
  {Sumithrarachchi}}]{MagneticFieldShift}%
  \BibitemOpen
  \bibfield  {author} {\bibinfo {author} {\bibfnamefont {R.}~\bibnamefont
  {Ringle}}, \bibinfo {author} {\bibfnamefont {T.}~\bibnamefont {Sun}},
  \bibinfo {author} {\bibfnamefont {G.}~\bibnamefont {Bollen}}, \bibinfo
  {author} {\bibfnamefont {D.}~\bibnamefont {Davies}}, \bibinfo {author}
  {\bibfnamefont {M.}~\bibnamefont {Facina}}, \bibinfo {author} {\bibfnamefont
  {J.}~\bibnamefont {Huikari}}, \bibinfo {author} {\bibfnamefont
  {E.}~\bibnamefont {Kwan}}, \bibinfo {author} {\bibfnamefont {D.~J.}\
  \bibnamefont {Morrissey}}, \bibinfo {author} {\bibfnamefont {A.}~\bibnamefont
  {Prinke}}, \bibinfo {author} {\bibfnamefont {J.}~\bibnamefont {Savory}},
  \bibinfo {author} {\bibfnamefont {P.}~\bibnamefont {Schury}}, \bibinfo
  {author} {\bibfnamefont {S.}~\bibnamefont {Schwarz}}, \ and\ \bibinfo
  {author} {\bibfnamefont {C.~S.}\ \bibnamefont {Sumithrarachchi}},\ }\href
  {\doibase 10.1103/PhysRevC.75.055503} {\bibfield  {journal} {\bibinfo
  {journal} {Phys. Rev. C}\ }\textbf {\bibinfo {volume} {75}},\ \bibinfo
  {pages} {055503} (\bibinfo {year} {2007}{\natexlab{b}})}\BibitemShut
  {NoStop}%
\bibitem [{\citenamefont {Birge}(1932)}]{BirgeRatio}%
  \BibitemOpen
  \bibfield  {author} {\bibinfo {author} {\bibfnamefont {R.~T.}\ \bibnamefont
  {Birge}},\ }\href {\doibase 10.1103/PhysRev.40.207} {\bibfield  {journal}
  {\bibinfo  {journal} {Phys. Rev.}\ }\textbf {\bibinfo {volume} {40}},\
  \bibinfo {pages} {207} (\bibinfo {year} {1932})}\BibitemShut {NoStop}%
\bibitem [{\citenamefont {Benenson}\ \emph {et~al.}(1977)\citenamefont
  {Benenson}, \citenamefont {Mueller}, \citenamefont {Kashy}, \citenamefont
  {Nann},\ and\ \citenamefont {Robinson}}]{Benenson19771187}%
  \BibitemOpen
  \bibfield  {author} {\bibinfo {author} {\bibfnamefont {W.}~\bibnamefont
  {Benenson}}, \bibinfo {author} {\bibfnamefont {D.}~\bibnamefont {Mueller}},
  \bibinfo {author} {\bibfnamefont {E.}~\bibnamefont {Kashy}}, \bibinfo
  {author} {\bibfnamefont {H.}~\bibnamefont {Nann}}, \ and\ \bibinfo {author}
  {\bibfnamefont {L.}~\bibnamefont {Robinson}},\ }\href {\doibase
  10.1103/PhysRevC.15.1187} {\bibfield  {journal} {\bibinfo  {journal}
  {Physical Review C}\ }\textbf {\bibinfo {volume} {15}},\ \bibinfo {pages}
  {1187 – 1190} (\bibinfo {year} {1977})},\ \bibinfo {note} {cited by:
  31}\BibitemShut {NoStop}%
\bibitem [{\citenamefont {Janiak}\ \emph {et~al.}(2017)\citenamefont {Janiak},
  \citenamefont {Soko\l{}owska}, \citenamefont {Bezbakh}, \citenamefont
  {Ciemny}, \citenamefont {Czyrkowski}, \citenamefont {Dabrowski},
  \citenamefont {Dominik}, \citenamefont {Fomichev}, \citenamefont {Golovkov},
  \citenamefont {Gorshkov}, \citenamefont {Janas}, \citenamefont
  {Kami\ifmmode~\acute{n}\else \'{n}\fi{}ski}, \citenamefont {Knyazev},
  \citenamefont {Krupko}, \citenamefont {Kuich}, \citenamefont {Mazzocchi},
  \citenamefont {Mentel}, \citenamefont {Pf\"utzner}, \citenamefont
  {Pluci\ifmmode~\acute{n}\else \'{n}\fi{}ski}, \citenamefont {Pomorski},
  \citenamefont {Slepniev},\ and\ \citenamefont
  {Zalewski}}]{PhysRevC.95.034315}%
  \BibitemOpen
  \bibfield  {author} {\bibinfo {author} {\bibfnamefont {L.}~\bibnamefont
  {Janiak}}, \bibinfo {author} {\bibfnamefont {N.}~\bibnamefont
  {Soko\l{}owska}}, \bibinfo {author} {\bibfnamefont {A.~A.}\ \bibnamefont
  {Bezbakh}}, \bibinfo {author} {\bibfnamefont {A.~A.}\ \bibnamefont {Ciemny}},
  \bibinfo {author} {\bibfnamefont {H.}~\bibnamefont {Czyrkowski}}, \bibinfo
  {author} {\bibfnamefont {R.}~\bibnamefont {Dabrowski}}, \bibinfo {author}
  {\bibfnamefont {W.}~\bibnamefont {Dominik}}, \bibinfo {author} {\bibfnamefont
  {A.~S.}\ \bibnamefont {Fomichev}}, \bibinfo {author} {\bibfnamefont {M.~S.}\
  \bibnamefont {Golovkov}}, \bibinfo {author} {\bibfnamefont {A.~V.}\
  \bibnamefont {Gorshkov}}, \bibinfo {author} {\bibfnamefont {Z.}~\bibnamefont
  {Janas}}, \bibinfo {author} {\bibfnamefont {G.}~\bibnamefont
  {Kami\ifmmode~\acute{n}\else \'{n}\fi{}ski}}, \bibinfo {author}
  {\bibfnamefont {A.~G.}\ \bibnamefont {Knyazev}}, \bibinfo {author}
  {\bibfnamefont {S.~A.}\ \bibnamefont {Krupko}}, \bibinfo {author}
  {\bibfnamefont {M.}~\bibnamefont {Kuich}}, \bibinfo {author} {\bibfnamefont
  {C.}~\bibnamefont {Mazzocchi}}, \bibinfo {author} {\bibfnamefont
  {M.}~\bibnamefont {Mentel}}, \bibinfo {author} {\bibfnamefont
  {M.}~\bibnamefont {Pf\"utzner}}, \bibinfo {author} {\bibfnamefont
  {P.}~\bibnamefont {Pluci\ifmmode~\acute{n}\else \'{n}\fi{}ski}}, \bibinfo
  {author} {\bibfnamefont {M.}~\bibnamefont {Pomorski}}, \bibinfo {author}
  {\bibfnamefont {R.~S.}\ \bibnamefont {Slepniev}}, \ and\ \bibinfo {author}
  {\bibfnamefont {B.}~\bibnamefont {Zalewski}},\ }\href {\doibase
  10.1103/PhysRevC.95.034315} {\bibfield  {journal} {\bibinfo  {journal} {Phys.
  Rev. C}\ }\textbf {\bibinfo {volume} {95}},\ \bibinfo {pages} {034315}
  (\bibinfo {year} {2017})}\BibitemShut {NoStop}%
\bibitem [{\citenamefont {Fu}\ \emph {et~al.}(2018)\citenamefont {Fu},
  \citenamefont {Zhang}, \citenamefont {Zhou}, \citenamefont {Wang},
  \citenamefont {Litvinov}, \citenamefont {Blaum}, \citenamefont {Xu},
  \citenamefont {Xu}, \citenamefont {Shuai}, \citenamefont {Lam}, \citenamefont
  {Chen}, \citenamefont {Yan}, \citenamefont {Bao}, \citenamefont {Chen},
  \citenamefont {Chen}, \citenamefont {He}, \citenamefont {Kubono},
  \citenamefont {Liu}, \citenamefont {Mao}, \citenamefont {Ma}, \citenamefont
  {Sun}, \citenamefont {Tu}, \citenamefont {Xing}, \citenamefont {Zhang},
  \citenamefont {Zeng}, \citenamefont {Zhou}, \citenamefont {Zhan},
  \citenamefont {Litvinov}, \citenamefont {Audi}, \citenamefont {Uesaka},
  \citenamefont {Yamaguchi}, \citenamefont {Yamaguchi}, \citenamefont {Ozawa},
  \citenamefont {Sun}, \citenamefont {Sun},\ and\ \citenamefont {Xu}}]{Fu2018}%
  \BibitemOpen
  \bibfield  {author} {\bibinfo {author} {\bibfnamefont {C.~Y.}\ \bibnamefont
  {Fu}}, \bibinfo {author} {\bibfnamefont {Y.~H.}\ \bibnamefont {Zhang}},
  \bibinfo {author} {\bibfnamefont {X.~H.}\ \bibnamefont {Zhou}}, \bibinfo
  {author} {\bibfnamefont {M.}~\bibnamefont {Wang}}, \bibinfo {author}
  {\bibfnamefont {Y.~A.}\ \bibnamefont {Litvinov}}, \bibinfo {author}
  {\bibfnamefont {K.}~\bibnamefont {Blaum}}, \bibinfo {author} {\bibfnamefont
  {H.~S.}\ \bibnamefont {Xu}}, \bibinfo {author} {\bibfnamefont
  {X.}~\bibnamefont {Xu}}, \bibinfo {author} {\bibfnamefont {P.}~\bibnamefont
  {Shuai}}, \bibinfo {author} {\bibfnamefont {Y.~H.}\ \bibnamefont {Lam}},
  \bibinfo {author} {\bibfnamefont {R.~J.}\ \bibnamefont {Chen}}, \bibinfo
  {author} {\bibfnamefont {X.~L.}\ \bibnamefont {Yan}}, \bibinfo {author}
  {\bibfnamefont {T.}~\bibnamefont {Bao}}, \bibinfo {author} {\bibfnamefont
  {X.~C.}\ \bibnamefont {Chen}}, \bibinfo {author} {\bibfnamefont
  {H.}~\bibnamefont {Chen}}, \bibinfo {author} {\bibfnamefont {J.~J.}\
  \bibnamefont {He}}, \bibinfo {author} {\bibfnamefont {S.}~\bibnamefont
  {Kubono}}, \bibinfo {author} {\bibfnamefont {D.~W.}\ \bibnamefont {Liu}},
  \bibinfo {author} {\bibfnamefont {R.~S.}\ \bibnamefont {Mao}}, \bibinfo
  {author} {\bibfnamefont {X.~W.}\ \bibnamefont {Ma}}, \bibinfo {author}
  {\bibfnamefont {M.~Z.}\ \bibnamefont {Sun}}, \bibinfo {author} {\bibfnamefont
  {X.~L.}\ \bibnamefont {Tu}}, \bibinfo {author} {\bibfnamefont {Y.~M.}\
  \bibnamefont {Xing}}, \bibinfo {author} {\bibfnamefont {P.}~\bibnamefont
  {Zhang}}, \bibinfo {author} {\bibfnamefont {Q.}~\bibnamefont {Zeng}},
  \bibinfo {author} {\bibfnamefont {X.}~\bibnamefont {Zhou}}, \bibinfo {author}
  {\bibfnamefont {W.~L.}\ \bibnamefont {Zhan}}, \bibinfo {author}
  {\bibfnamefont {S.}~\bibnamefont {Litvinov}}, \bibinfo {author}
  {\bibfnamefont {G.}~\bibnamefont {Audi}}, \bibinfo {author} {\bibfnamefont
  {T.}~\bibnamefont {Uesaka}}, \bibinfo {author} {\bibfnamefont
  {Y.}~\bibnamefont {Yamaguchi}}, \bibinfo {author} {\bibfnamefont
  {T.}~\bibnamefont {Yamaguchi}}, \bibinfo {author} {\bibfnamefont
  {A.}~\bibnamefont {Ozawa}}, \bibinfo {author} {\bibfnamefont {B.~H.}\
  \bibnamefont {Sun}}, \bibinfo {author} {\bibfnamefont {Y.}~\bibnamefont
  {Sun}}, \ and\ \bibinfo {author} {\bibfnamefont {F.~R.}\ \bibnamefont {Xu}},\
  }\href {\doibase 10.1103/PhysRevC.98.014315} {\bibfield  {journal} {\bibinfo
  {journal} {Phys. Rev. C}\ }\textbf {\bibinfo {volume} {98}},\ \bibinfo
  {pages} {014315} (\bibinfo {year} {2018})}\BibitemShut {NoStop}%
\bibitem [{\citenamefont {Sun}\ \emph {et~al.}(2020)\citenamefont {Sun},
  \citenamefont {Xu}, \citenamefont {Hou}, \citenamefont {Lin}, \citenamefont
  {José}, \citenamefont {Lee}, \citenamefont {He}, \citenamefont {Li},
  \citenamefont {Wang}, \citenamefont {Yuan}, \citenamefont {Herwig},
  \citenamefont {Keegans}, \citenamefont {Budner}, \citenamefont {Wang},
  \citenamefont {Wu}, \citenamefont {Liang}, \citenamefont {Yang},
  \citenamefont {Lam}, \citenamefont {Ma}, \citenamefont {Duan}, \citenamefont
  {Gao}, \citenamefont {Hu}, \citenamefont {Bai}, \citenamefont {Ma},
  \citenamefont {Wang}, \citenamefont {Zhong}, \citenamefont {Wu},
  \citenamefont {Luo}, \citenamefont {Jiang}, \citenamefont {Liu},
  \citenamefont {Hou}, \citenamefont {Li}, \citenamefont {Ma}, \citenamefont
  {Ma}, \citenamefont {Shi}, \citenamefont {Yu}, \citenamefont {Patel},
  \citenamefont {Jin}, \citenamefont {Wang}, \citenamefont {Yu}, \citenamefont
  {Zhou}, \citenamefont {Wang}, \citenamefont {Hu}, \citenamefont {Wang},
  \citenamefont {Zang}, \citenamefont {Li}, \citenamefont {Zhao}, \citenamefont
  {Jia}, \citenamefont {Yang}, \citenamefont {Wen}, \citenamefont {Yang},
  \citenamefont {Pan}, \citenamefont {Wang}, \citenamefont {Hu}, \citenamefont
  {Chen}, \citenamefont {Liu}, \citenamefont {Yang},\ and\ \citenamefont
  {Zhao}}]{LJSunMeasurement}%
  \BibitemOpen
  \bibfield  {author} {\bibinfo {author} {\bibfnamefont {L.}~\bibnamefont
  {Sun}}, \bibinfo {author} {\bibfnamefont {X.}~\bibnamefont {Xu}}, \bibinfo
  {author} {\bibfnamefont {S.}~\bibnamefont {Hou}}, \bibinfo {author}
  {\bibfnamefont {C.}~\bibnamefont {Lin}}, \bibinfo {author} {\bibfnamefont
  {J.}~\bibnamefont {José}}, \bibinfo {author} {\bibfnamefont
  {J.}~\bibnamefont {Lee}}, \bibinfo {author} {\bibfnamefont {J.}~\bibnamefont
  {He}}, \bibinfo {author} {\bibfnamefont {Z.}~\bibnamefont {Li}}, \bibinfo
  {author} {\bibfnamefont {J.}~\bibnamefont {Wang}}, \bibinfo {author}
  {\bibfnamefont {C.}~\bibnamefont {Yuan}}, \bibinfo {author} {\bibfnamefont
  {F.}~\bibnamefont {Herwig}}, \bibinfo {author} {\bibfnamefont
  {J.}~\bibnamefont {Keegans}}, \bibinfo {author} {\bibfnamefont
  {T.}~\bibnamefont {Budner}}, \bibinfo {author} {\bibfnamefont
  {D.}~\bibnamefont {Wang}}, \bibinfo {author} {\bibfnamefont {H.}~\bibnamefont
  {Wu}}, \bibinfo {author} {\bibfnamefont {P.}~\bibnamefont {Liang}}, \bibinfo
  {author} {\bibfnamefont {Y.}~\bibnamefont {Yang}}, \bibinfo {author}
  {\bibfnamefont {Y.}~\bibnamefont {Lam}}, \bibinfo {author} {\bibfnamefont
  {P.}~\bibnamefont {Ma}}, \bibinfo {author} {\bibfnamefont {F.}~\bibnamefont
  {Duan}}, \bibinfo {author} {\bibfnamefont {Z.}~\bibnamefont {Gao}}, \bibinfo
  {author} {\bibfnamefont {Q.}~\bibnamefont {Hu}}, \bibinfo {author}
  {\bibfnamefont {Z.}~\bibnamefont {Bai}}, \bibinfo {author} {\bibfnamefont
  {J.}~\bibnamefont {Ma}}, \bibinfo {author} {\bibfnamefont {J.}~\bibnamefont
  {Wang}}, \bibinfo {author} {\bibfnamefont {F.}~\bibnamefont {Zhong}},
  \bibinfo {author} {\bibfnamefont {C.}~\bibnamefont {Wu}}, \bibinfo {author}
  {\bibfnamefont {D.}~\bibnamefont {Luo}}, \bibinfo {author} {\bibfnamefont
  {Y.}~\bibnamefont {Jiang}}, \bibinfo {author} {\bibfnamefont
  {Y.}~\bibnamefont {Liu}}, \bibinfo {author} {\bibfnamefont {D.}~\bibnamefont
  {Hou}}, \bibinfo {author} {\bibfnamefont {R.}~\bibnamefont {Li}}, \bibinfo
  {author} {\bibfnamefont {N.}~\bibnamefont {Ma}}, \bibinfo {author}
  {\bibfnamefont {W.}~\bibnamefont {Ma}}, \bibinfo {author} {\bibfnamefont
  {G.}~\bibnamefont {Shi}}, \bibinfo {author} {\bibfnamefont {G.}~\bibnamefont
  {Yu}}, \bibinfo {author} {\bibfnamefont {D.}~\bibnamefont {Patel}}, \bibinfo
  {author} {\bibfnamefont {S.}~\bibnamefont {Jin}}, \bibinfo {author}
  {\bibfnamefont {Y.}~\bibnamefont {Wang}}, \bibinfo {author} {\bibfnamefont
  {Y.}~\bibnamefont {Yu}}, \bibinfo {author} {\bibfnamefont {Q.}~\bibnamefont
  {Zhou}}, \bibinfo {author} {\bibfnamefont {P.}~\bibnamefont {Wang}}, \bibinfo
  {author} {\bibfnamefont {L.}~\bibnamefont {Hu}}, \bibinfo {author}
  {\bibfnamefont {X.}~\bibnamefont {Wang}}, \bibinfo {author} {\bibfnamefont
  {H.}~\bibnamefont {Zang}}, \bibinfo {author} {\bibfnamefont {P.}~\bibnamefont
  {Li}}, \bibinfo {author} {\bibfnamefont {Q.}~\bibnamefont {Zhao}}, \bibinfo
  {author} {\bibfnamefont {H.}~\bibnamefont {Jia}}, \bibinfo {author}
  {\bibfnamefont {L.}~\bibnamefont {Yang}}, \bibinfo {author} {\bibfnamefont
  {P.}~\bibnamefont {Wen}}, \bibinfo {author} {\bibfnamefont {F.}~\bibnamefont
  {Yang}}, \bibinfo {author} {\bibfnamefont {M.}~\bibnamefont {Pan}}, \bibinfo
  {author} {\bibfnamefont {X.}~\bibnamefont {Wang}}, \bibinfo {author}
  {\bibfnamefont {Z.}~\bibnamefont {Hu}}, \bibinfo {author} {\bibfnamefont
  {R.}~\bibnamefont {Chen}}, \bibinfo {author} {\bibfnamefont {M.}~\bibnamefont
  {Liu}}, \bibinfo {author} {\bibfnamefont {W.}~\bibnamefont {Yang}}, \ and\
  \bibinfo {author} {\bibfnamefont {Y.}~\bibnamefont {Zhao}},\ }\href {\doibase
  https://doi.org/10.1016/j.physletb.2020.135213} {\bibfield  {journal}
  {\bibinfo  {journal} {Physics Letters B}\ }\textbf {\bibinfo {volume}
  {802}},\ \bibinfo {pages} {135213} (\bibinfo {year} {2020})}\BibitemShut
  {NoStop}%
\bibitem [{\citenamefont {Wang}\ \emph {et~al.}(2012)\citenamefont {Wang},
  \citenamefont {Audi}, \citenamefont {Wapstra}, \citenamefont {Kondev},
  \citenamefont {MacCormick}, \citenamefont {Xu},\ and\ \citenamefont
  {Pfeiffer}}]{AME2012}%
  \BibitemOpen
  \bibfield  {author} {\bibinfo {author} {\bibfnamefont {M.}~\bibnamefont
  {Wang}}, \bibinfo {author} {\bibfnamefont {G.}~\bibnamefont {Audi}}, \bibinfo
  {author} {\bibfnamefont {A.}~\bibnamefont {Wapstra}}, \bibinfo {author}
  {\bibfnamefont {F.}~\bibnamefont {Kondev}}, \bibinfo {author} {\bibfnamefont
  {M.}~\bibnamefont {MacCormick}}, \bibinfo {author} {\bibfnamefont
  {X.}~\bibnamefont {Xu}}, \ and\ \bibinfo {author} {\bibfnamefont
  {B.}~\bibnamefont {Pfeiffer}},\ }\href {\doibase 10.1088/1674-1137/36/12/003}
  {\bibfield  {journal} {\bibinfo  {journal} {Chinese Physics C}\ }\textbf
  {\bibinfo {volume} {36}},\ \bibinfo {pages} {1603} (\bibinfo {year}
  {2012})}\BibitemShut {NoStop}%
\bibitem [{\citenamefont {Wang}\ \emph {et~al.}(2021)\citenamefont {Wang},
  \citenamefont {Huang}, \citenamefont {Kondev}, \citenamefont {Audi},\ and\
  \citenamefont {Naimi}}]{AME2020}%
  \BibitemOpen
  \bibfield  {author} {\bibinfo {author} {\bibfnamefont {M.}~\bibnamefont
  {Wang}}, \bibinfo {author} {\bibfnamefont {W.}~\bibnamefont {Huang}},
  \bibinfo {author} {\bibfnamefont {F.}~\bibnamefont {Kondev}}, \bibinfo
  {author} {\bibfnamefont {G.}~\bibnamefont {Audi}}, \ and\ \bibinfo {author}
  {\bibfnamefont {S.}~\bibnamefont {Naimi}},\ }\href {\doibase
  10.1088/1674-1137/abddaf} {\bibfield  {journal} {\bibinfo  {journal} {Chinese
  Physics C}\ }\textbf {\bibinfo {volume} {45}},\ \bibinfo {pages} {030003}
  (\bibinfo {year} {2021})}\BibitemShut {NoStop}%
\bibitem [{\citenamefont {{Barker}}\ \emph {et~al.}(1971)\citenamefont
  {{Barker}}, \citenamefont {{Carter}}, \citenamefont {{Kean}}, \citenamefont
  {{Piluso}},\ and\ \citenamefont {{Spear}}}]{OldSi27}%
  \BibitemOpen
  \bibfield  {author} {\bibinfo {author} {\bibfnamefont {F.~C.}\ \bibnamefont
  {{Barker}}}, \bibinfo {author} {\bibfnamefont {K.~W.}\ \bibnamefont
  {{Carter}}}, \bibinfo {author} {\bibfnamefont {D.~C.}\ \bibnamefont
  {{Kean}}}, \bibinfo {author} {\bibfnamefont {C.~J.}\ \bibnamefont
  {{Piluso}}}, \ and\ \bibinfo {author} {\bibfnamefont {R.~H.}\ \bibnamefont
  {{Spear}}},\ }\href@noop {} {\bibfield  {journal} {\bibinfo  {journal}
  {Aust.J.Phys.}\ }\textbf {\bibinfo {volume} {24}},\ \bibinfo {pages} {1}
  (\bibinfo {year} {1971})}\BibitemShut {NoStop}%
\bibitem [{\citenamefont {McCleskey}\ \emph {et~al.}(2016)\citenamefont
  {McCleskey}, \citenamefont {Banu}, \citenamefont {McCleskey}, \citenamefont
  {Davinson}, \citenamefont {Doherty}, \citenamefont {Lotay}, \citenamefont
  {Roeder}, \citenamefont {Saastamoinen}, \citenamefont {Spiridon},
  \citenamefont {Trache}, \citenamefont {Wallace}, \citenamefont {Woods},\ and\
  \citenamefont {Tribble}}]{McCleskey2016}%
  \BibitemOpen
  \bibfield  {author} {\bibinfo {author} {\bibfnamefont {E.}~\bibnamefont
  {McCleskey}}, \bibinfo {author} {\bibfnamefont {A.}~\bibnamefont {Banu}},
  \bibinfo {author} {\bibfnamefont {M.}~\bibnamefont {McCleskey}}, \bibinfo
  {author} {\bibfnamefont {T.}~\bibnamefont {Davinson}}, \bibinfo {author}
  {\bibfnamefont {D.~T.}\ \bibnamefont {Doherty}}, \bibinfo {author}
  {\bibfnamefont {G.}~\bibnamefont {Lotay}}, \bibinfo {author} {\bibfnamefont
  {B.~T.}\ \bibnamefont {Roeder}}, \bibinfo {author} {\bibfnamefont
  {A.}~\bibnamefont {Saastamoinen}}, \bibinfo {author} {\bibfnamefont
  {A.}~\bibnamefont {Spiridon}}, \bibinfo {author} {\bibfnamefont
  {L.}~\bibnamefont {Trache}}, \bibinfo {author} {\bibfnamefont {J.~P.}\
  \bibnamefont {Wallace}}, \bibinfo {author} {\bibfnamefont {P.~J.}\
  \bibnamefont {Woods}}, \ and\ \bibinfo {author} {\bibfnamefont {R.~E.}\
  \bibnamefont {Tribble}},\ }\href {\doibase 10.1103/PhysRevC.94.065806}
  {\bibfield  {journal} {\bibinfo  {journal} {Phys. Rev. C}\ }\textbf {\bibinfo
  {volume} {94}},\ \bibinfo {pages} {065806} (\bibinfo {year}
  {2016})}\BibitemShut {NoStop}%
\bibitem [{\citenamefont {Basunia}(2010)}]{Al27}%
  \BibitemOpen
  \bibfield  {author} {\bibinfo {author} {\bibfnamefont {M.~S.}\ \bibnamefont
  {Basunia}},\ }\href {https://www.nndc.bnl.gov/ensnds/27/Al/26mg_p_g.pdf}
  {\bibfield  {journal} {\bibinfo  {journal} {Nucl. Data Sheets}\ }\textbf
  {\bibinfo {volume} {112}},\ \bibinfo {pages} {1875} (\bibinfo {year}
  {2010})}\BibitemShut {NoStop}%
\bibitem [{\citenamefont {Sun}\ \emph {et~al.}(2019)\citenamefont {Sun},
  \citenamefont {Xu}, \citenamefont {Lin}, \citenamefont {Lee}, \citenamefont
  {Hou}, \citenamefont {Yuan}, \citenamefont {Li}, \citenamefont {Jos{\'{e} }},
  \citenamefont {He}, \citenamefont {Wang}, \citenamefont {Wang}, \citenamefont
  {Wu}, \citenamefont {Liang}, \citenamefont {Yang}, \citenamefont {Lam},
  \citenamefont {Ma}, \citenamefont {Duan}, \citenamefont {Gao}, \citenamefont
  {Hu}, \citenamefont {Bai}, \citenamefont {Ma}, \citenamefont {Wang},
  \citenamefont {Zhong}, \citenamefont {Wu}, \citenamefont {Luo}, \citenamefont
  {Jiang}, \citenamefont {Liu}, \citenamefont {Hou}, \citenamefont {Li},
  \citenamefont {Ma}, \citenamefont {Ma}, \citenamefont {Shi}, \citenamefont
  {Yu}, \citenamefont {Patel}, \citenamefont {Jin}, \citenamefont {Wang},
  \citenamefont {Yu}, \citenamefont {Zhou}, \citenamefont {Wang}, \citenamefont
  {Hu}, \citenamefont {Wang}, \citenamefont {Zang}, \citenamefont {Li},
  \citenamefont {Zhao}, \citenamefont {Yang}, \citenamefont {Wen},
  \citenamefont {Yang}, \citenamefont {Jia}, \citenamefont {Zhang},
  \citenamefont {Pan}, \citenamefont {Wang}, \citenamefont {Sun}, \citenamefont
  {Hu}, \citenamefont {Chen}, \citenamefont {Liu}, \citenamefont {Yang},
  \citenamefont {Zhao},\ and\ \citenamefont {and}}]{Sun_2019}%
  \BibitemOpen
  \bibfield  {author} {\bibinfo {author} {\bibfnamefont {L.~J.}\ \bibnamefont
  {Sun}}, \bibinfo {author} {\bibfnamefont {X.~X.}\ \bibnamefont {Xu}},
  \bibinfo {author} {\bibfnamefont {C.~J.}\ \bibnamefont {Lin}}, \bibinfo
  {author} {\bibfnamefont {J.}~\bibnamefont {Lee}}, \bibinfo {author}
  {\bibfnamefont {S.~Q.}\ \bibnamefont {Hou}}, \bibinfo {author} {\bibfnamefont
  {C.~X.}\ \bibnamefont {Yuan}}, \bibinfo {author} {\bibfnamefont {Z.~H.}\
  \bibnamefont {Li}}, \bibinfo {author} {\bibfnamefont {J.}~\bibnamefont
  {Jos{\'{e} }}}, \bibinfo {author} {\bibfnamefont {J.~J.}\ \bibnamefont {He}},
  \bibinfo {author} {\bibfnamefont {J.~S.}\ \bibnamefont {Wang}}, \bibinfo
  {author} {\bibfnamefont {D.~X.}\ \bibnamefont {Wang}}, \bibinfo {author}
  {\bibfnamefont {H.~Y.}\ \bibnamefont {Wu}}, \bibinfo {author} {\bibfnamefont
  {P.~F.}\ \bibnamefont {Liang}}, \bibinfo {author} {\bibfnamefont {Y.~Y.}\
  \bibnamefont {Yang}}, \bibinfo {author} {\bibfnamefont {Y.~H.}\ \bibnamefont
  {Lam}}, \bibinfo {author} {\bibfnamefont {P.}~\bibnamefont {Ma}}, \bibinfo
  {author} {\bibfnamefont {F.~F.}\ \bibnamefont {Duan}}, \bibinfo {author}
  {\bibfnamefont {Z.~H.}\ \bibnamefont {Gao}}, \bibinfo {author} {\bibfnamefont
  {Q.}~\bibnamefont {Hu}}, \bibinfo {author} {\bibfnamefont {Z.}~\bibnamefont
  {Bai}}, \bibinfo {author} {\bibfnamefont {J.~B.}\ \bibnamefont {Ma}},
  \bibinfo {author} {\bibfnamefont {J.~G.}\ \bibnamefont {Wang}}, \bibinfo
  {author} {\bibfnamefont {F.~P.}\ \bibnamefont {Zhong}}, \bibinfo {author}
  {\bibfnamefont {C.~G.}\ \bibnamefont {Wu}}, \bibinfo {author} {\bibfnamefont
  {D.~W.}\ \bibnamefont {Luo}}, \bibinfo {author} {\bibfnamefont
  {Y.}~\bibnamefont {Jiang}}, \bibinfo {author} {\bibfnamefont
  {Y.}~\bibnamefont {Liu}}, \bibinfo {author} {\bibfnamefont {D.~S.}\
  \bibnamefont {Hou}}, \bibinfo {author} {\bibfnamefont {R.}~\bibnamefont
  {Li}}, \bibinfo {author} {\bibfnamefont {N.~R.}\ \bibnamefont {Ma}}, \bibinfo
  {author} {\bibfnamefont {W.~H.}\ \bibnamefont {Ma}}, \bibinfo {author}
  {\bibfnamefont {G.~Z.}\ \bibnamefont {Shi}}, \bibinfo {author} {\bibfnamefont
  {G.~M.}\ \bibnamefont {Yu}}, \bibinfo {author} {\bibfnamefont
  {D.}~\bibnamefont {Patel}}, \bibinfo {author} {\bibfnamefont {S.~Y.}\
  \bibnamefont {Jin}}, \bibinfo {author} {\bibfnamefont {Y.~F.}\ \bibnamefont
  {Wang}}, \bibinfo {author} {\bibfnamefont {Y.~C.}\ \bibnamefont {Yu}},
  \bibinfo {author} {\bibfnamefont {Q.~W.}\ \bibnamefont {Zhou}}, \bibinfo
  {author} {\bibfnamefont {P.}~\bibnamefont {Wang}}, \bibinfo {author}
  {\bibfnamefont {L.~Y.}\ \bibnamefont {Hu}}, \bibinfo {author} {\bibfnamefont
  {X.}~\bibnamefont {Wang}}, \bibinfo {author} {\bibfnamefont {H.~L.}\
  \bibnamefont {Zang}}, \bibinfo {author} {\bibfnamefont {P.~J.}\ \bibnamefont
  {Li}}, \bibinfo {author} {\bibfnamefont {Q.~Q.}\ \bibnamefont {Zhao}},
  \bibinfo {author} {\bibfnamefont {L.}~\bibnamefont {Yang}}, \bibinfo {author}
  {\bibfnamefont {P.~W.}\ \bibnamefont {Wen}}, \bibinfo {author} {\bibfnamefont
  {F.}~\bibnamefont {Yang}}, \bibinfo {author} {\bibfnamefont {H.~M.}\
  \bibnamefont {Jia}}, \bibinfo {author} {\bibfnamefont {G.~L.}\ \bibnamefont
  {Zhang}}, \bibinfo {author} {\bibfnamefont {M.}~\bibnamefont {Pan}}, \bibinfo
  {author} {\bibfnamefont {X.~Y.}\ \bibnamefont {Wang}}, \bibinfo {author}
  {\bibfnamefont {H.~H.}\ \bibnamefont {Sun}}, \bibinfo {author} {\bibfnamefont
  {Z.~G.}\ \bibnamefont {Hu}}, \bibinfo {author} {\bibfnamefont {R.~F.}\
  \bibnamefont {Chen}}, \bibinfo {author} {\bibfnamefont {M.~L.}\ \bibnamefont
  {Liu}}, \bibinfo {author} {\bibfnamefont {W.~Q.}\ \bibnamefont {Yang}},
  \bibinfo {author} {\bibfnamefont {Y.~M.}\ \bibnamefont {Zhao}}, \ and\
  \bibinfo {author} {\bibfnamefont {H.~Q.~Z.}\ \bibnamefont {and}},\ }\href
  {\doibase 10.1103/physrevc.99.064312} {\bibfield  {journal} {\bibinfo
  {journal} {Physical Review C}\ }\textbf {\bibinfo {volume} {99}} (\bibinfo
  {year} {2019}),\ 10.1103/physrevc.99.064312}\BibitemShut {NoStop}%
\bibitem [{\citenamefont {Schatz}\ \emph {et~al.}(1998)\citenamefont {Schatz},
  \citenamefont {Aprahamian}, \citenamefont {Görres}, \citenamefont
  {Wiescher}, \citenamefont {Rauscher}, \citenamefont {Rembges}, \citenamefont
  {Thielemann}, \citenamefont {Pfeiffer}, \citenamefont {Möller},
  \citenamefont {Kratz}, \citenamefont {Herndl}, \citenamefont {Brown},\ and\
  \citenamefont {Rebel}}]{ReverseRate}%
  \BibitemOpen
  \bibfield  {author} {\bibinfo {author} {\bibfnamefont {H.}~\bibnamefont
  {Schatz}}, \bibinfo {author} {\bibfnamefont {A.}~\bibnamefont {Aprahamian}},
  \bibinfo {author} {\bibfnamefont {J.}~\bibnamefont {Görres}}, \bibinfo
  {author} {\bibfnamefont {M.}~\bibnamefont {Wiescher}}, \bibinfo {author}
  {\bibfnamefont {T.}~\bibnamefont {Rauscher}}, \bibinfo {author}
  {\bibfnamefont {J.}~\bibnamefont {Rembges}}, \bibinfo {author} {\bibfnamefont
  {F.-K.}\ \bibnamefont {Thielemann}}, \bibinfo {author} {\bibfnamefont
  {B.}~\bibnamefont {Pfeiffer}}, \bibinfo {author} {\bibfnamefont
  {P.}~\bibnamefont {Möller}}, \bibinfo {author} {\bibfnamefont {K.-L.}\
  \bibnamefont {Kratz}}, \bibinfo {author} {\bibfnamefont {H.}~\bibnamefont
  {Herndl}}, \bibinfo {author} {\bibfnamefont {B.}~\bibnamefont {Brown}}, \
  and\ \bibinfo {author} {\bibfnamefont {H.}~\bibnamefont {Rebel}},\ }\href
  {\doibase https://doi.org/10.1016/S0370-1573(97)00048-3} {\bibfield
  {journal} {\bibinfo  {journal} {Physics Reports}\ }\textbf {\bibinfo {volume}
  {294}},\ \bibinfo {pages} {167} (\bibinfo {year} {1998})}\BibitemShut
  {NoStop}%
\bibitem [{\citenamefont {Meisel}(2018)}]{Meisel_2018}%
  \BibitemOpen
  \bibfield  {author} {\bibinfo {author} {\bibfnamefont {Z.}~\bibnamefont
  {Meisel}},\ }\href {\doibase 10.3847/1538-4357/aac3d3} {\bibfield  {journal}
  {\bibinfo  {journal} {The Astrophysical Journal}\ }\textbf {\bibinfo {volume}
  {860}},\ \bibinfo {pages} {147} (\bibinfo {year} {2018})}\BibitemShut
  {NoStop}%
\bibitem [{\citenamefont {Cyburt}\ \emph {et~al.}(2016)\citenamefont {Cyburt},
  \citenamefont {Amthor}, \citenamefont {Heger}, \citenamefont {Johnson},
  \citenamefont {Keek}, \citenamefont {Meisel}, \citenamefont {Schatz},\ and\
  \citenamefont {Smith}}]{Cyburt_2016}%
  \BibitemOpen
  \bibfield  {author} {\bibinfo {author} {\bibfnamefont {R.~H.}\ \bibnamefont
  {Cyburt}}, \bibinfo {author} {\bibfnamefont {A.~M.}\ \bibnamefont {Amthor}},
  \bibinfo {author} {\bibfnamefont {A.}~\bibnamefont {Heger}}, \bibinfo
  {author} {\bibfnamefont {E.}~\bibnamefont {Johnson}}, \bibinfo {author}
  {\bibfnamefont {L.}~\bibnamefont {Keek}}, \bibinfo {author} {\bibfnamefont
  {Z.}~\bibnamefont {Meisel}}, \bibinfo {author} {\bibfnamefont
  {H.}~\bibnamefont {Schatz}}, \ and\ \bibinfo {author} {\bibfnamefont
  {K.}~\bibnamefont {Smith}},\ }\href {\doibase 10.3847/0004-637X/830/2/55}
  {\bibfield  {journal} {\bibinfo  {journal} {The Astrophysical Journal}\
  }\textbf {\bibinfo {volume} {830}},\ \bibinfo {pages} {55} (\bibinfo {year}
  {2016})}\BibitemShut {NoStop}%
\bibitem [{\citenamefont {Rauscher}\ \emph {et~al.}(2016)\citenamefont
  {Rauscher}, \citenamefont {Nishimura}, \citenamefont {Hirschi}, \citenamefont
  {Cescutti}, \citenamefont {Murphy},\ and\ \citenamefont
  {Heger}}]{NuclearDataUncertainties}%
  \BibitemOpen
  \bibfield  {author} {\bibinfo {author} {\bibfnamefont {T.}~\bibnamefont
  {Rauscher}}, \bibinfo {author} {\bibfnamefont {N.}~\bibnamefont {Nishimura}},
  \bibinfo {author} {\bibfnamefont {R.}~\bibnamefont {Hirschi}}, \bibinfo
  {author} {\bibfnamefont {G.}~\bibnamefont {Cescutti}}, \bibinfo {author}
  {\bibfnamefont {A.~S.~J.}\ \bibnamefont {Murphy}}, \ and\ \bibinfo {author}
  {\bibfnamefont {A.}~\bibnamefont {Heger}},\ }\href {\doibase
  10.1093/mnras/stw2266} {\bibfield  {journal} {\bibinfo  {journal} {Monthly
  Notices of the Royal Astronomical Society}\ }\textbf {\bibinfo {volume}
  {463}},\ \bibinfo {pages} {4153} (\bibinfo {year} {2016})}\BibitemShut
  {NoStop}%
\end{thebibliography}%
\vspace{1mm}

\end{document}